\documentclass[%
 preprint,
groupedaddress,
superscriptaddress,
 amsmath,amssymb,
 aps,
 prx,
 showpacs
]{revtex4-2}

\usepackage{graphicx}
\usepackage{makecell,booktabs}
\usepackage{dcolumn}
\usepackage{bm}
\usepackage{hyperref}
\hypersetup{
    colorlinks=true,
    urlcolor= blue,
    citecolor=blue,
linkcolor= blue}
\usepackage{xcolor}

\newcommand{\BiSb}{$\mathrm{Bi_{1-x}Sb_x}$}

\renewcommand{\v}[1]{\ensuremath{\mathbf{#1}}}

\newcommand{\al}[1]{\begin{align}#1\end{align}}
\newcommand{\bs}{\begin{split}}
\newcommand{\es}{\end{split}}

\begin{document}


\title{Spin Hall conductivity in \BiSb~as an experimental test of bulk-boundary correspondence}

\author{Yongxi Ou}
\author{Wilson Yanez-Parre\~no}
\author{Yu-sheng Huang}

 \affiliation{Department of Physics, Pennsylvania State University, University Park, Pennsylvania 16802, USA}

\author{Supriya Ghosh}
\affiliation{%
Department of Chemical Engineering and Materials Science, University of Minnesota, Minneapolis, Minnesota 55455, USA
}%
\author{C\"uneyt \c{S}ahin}
\affiliation{UNAM — National Nanotechnology Research Center and Institute of Materials Science and Nanotechnology, Bilkent University, Ankara, Turkey}
\affiliation{Department of Physics, University of Iowa, Iowa City IA, USA}

\author{Max Stanley}
 \affiliation{Department of Physics, Pennsylvania State University, University Park, Pennsylvania 16802, USA}
 \author{Sandra Santhosh}
 \affiliation{Department of Physics, Pennsylvania State University, University Park, Pennsylvania 16802, USA}
 \author{Saurav Islam}
 \affiliation{Department of Physics, Pennsylvania State University, University Park, Pennsylvania 16802, USA}
 \author{Anthony Richardella}
 \affiliation{Department of Physics, Pennsylvania State University, University Park, Pennsylvania 16802, USA}
 \affiliation{Materials Research Institute, Pennsylvania State University, University Park, Pennsylvania 16802, USA}

\author{K. Andre Mkhoyan}
\affiliation{%
Department of Chemical Engineering and Materials Science, University of Minnesota, Minneapolis, Minnesota 55455, USA
}%

\author{Michael E. Flatt\'e}
\affiliation{Department of Physics and Astronomy, University of Iowa, Iowa City IA, USA}
\author{Nitin Samarth}
\email{nsamarth@psu.edu}
\affiliation{Department of Physics, Pennsylvania State University, University Park, Pennsylvania 16802, USA}
 \affiliation{Materials Research Institute, Pennsylvania State University, University Park, Pennsylvania 16802, USA}
  \affiliation{Department of Materials Science and Engineering, Pennsylvania State University, University Park, Pennsylvania 16802, USA}

\begin{abstract}
Bulk-boundary correspondence is a foundational principle underlying the electronic band structure and physical behavior of topological quantum materials. Although it has been rigorously tested in topological systems where the physical properties involve charge currents, it remains unclear whether bulk-boundary correspondence should also hold for non-conserved spin currents. We study charge-to-spin conversion in a canonical topological insulator, \BiSb, to address this fundamentally unresolved question. We use spin-torque ferromagnetic resonance measurements to accurately probe the charge-to-spin conversion efficiency in epitaxial \BiSb~thin films of high structural quality spanning the entire range of composition, including both trivial and topological band structures, as verified using {\it in vacuo} angle-resolved photoemission spectroscopy. From these measurements, we deduce the effective spin Hall conductivity (SHC) and find excellent agreement with the values predicted by tight-binding calculations for the intrinsic SHC of the bulk bands. These results provide strong evidence that the strong spin-orbit entanglement of bulk states well below the Fermi energy connects directly to the SHC in epitaxial \BiSb~films interfaced with a metallic ferromagnet. The excellent agreement between theory and experiment points to the generic value of analyses focused entirely on bulk properties, even for topological systems involving non-conserved spin currents. 

\end{abstract}

\date{\today}

\maketitle

\newpage

The bulk-boundary correspondence principle has played an important role in the fundamental understanding of physical problems in diverse contexts, ranging from quantum field theory~\cite{VBala_WOS:000079938100080} to topological quantum materials ~\cite{Hasan_RevModPhys.82.3045,Teo_PRB_2010,Schindler_NatPhys_2018}. In its simplest form it posits that a material with a nontrivial topological quantized invariant in a gapped bulk will produce topologically protected  states at the Fermi energy at appropriate edges, however the concept appears to have more general implications.  For example, in quantum Hall insulators the same Hall conductance results solely from a bulk picture (Streda formula) that sums the contributions of the ``Fermi sea'' states in a gapped material, and also solely from an edge state picture that considers ``Fermi energy'' states; when measured the division of the currents between bulk and edge is nonuniversal, although the sum is the quantized value, and there are also scenarios (Corbino geometry) where the bulk expression must be used. The bulk boundary correspondence has been confirmed for these systems\cite{KvK,Uri_WOS:000507726300002,Johnsen_WOS:000961165700005}, in topological semimetals~\cite{Xu_SciAdv_2015}, and complex current flow in quantum anomalous Hall insulators~\cite{Ferguson_2023} which also involve conserved charge currents. 
A fundamental question that arises is to what degree bulk-boundary correspondence exists or is useful for pure spin currents in a topological quantum material despite the absence of a quantized topological invariant and the lack of conservation of  spin. A calculation of the spin Hall conductivity (SHC) of \BiSb, based entirely on the Berry curvature in the bulk Fermi sea (resulting from strong spin-orbit coupling), predicted a large SHC over a broad range of composition, including both the topological insulator (TI) and trivial regimes~\cite{Sahin2015}, including an approximate continuity of the SHC across the topological/trivial composition boundary. This theoretical prediction and the underlying assumption (based on an expanded bulk-boundary correspondence hypothesis) that a solely bulk Fermi sea calculation would accurately reflect the SHC coming from bulk and edge state contributions together has not yet been rigorously tested in experiments. Indeed, experimental measurements of the spin-charge interconversion efficiency, often identified by the spin Hall angle ($\theta_{\textrm{SH}} $), in \BiSb~have shown large discrepancies amongst themselves (ranging over almost two orders of magnitude) ~\cite{Khang2018,Roschewsky2019,Khang2020, Chi2020}. This motivates additional rigorous experimental tests of the theoretical calculations. 

In this paper, we accurately measure the SHC in a TI/metallic ferromagnet heterostructure geometry (\BiSb/Ni$_{0.8}$Fe$_{0.2}$) by using spin torque ferromagnetic resonance (ST-FMR) to probe the generation of spin-orbit torque (SOT) by an electrical current ~\cite{PhysRevLett.106.036601LiuSTFMR}. The experiments use epitaxial \BiSb~thin films of good structural quality grown by molecular beam epitaxy (MBE), with the band structure thoroughly characterized via {\it in vacuo} angle resolved photoemission spectroscopy (ARPES). By systematically varying the alloy composition, we measure the SOT efficiency (directly related to $\theta_{\textrm{SH}} $) and deduce the SHC over the entire composition range from pure Bi to pure Sb. This enables us to demonstrate a rigorous agreement between theoretical calculations of the SHC using purely bulk states and experimental measurements of charge-spin conversion at the interface. Our results demonstrate the validity of bulk-boundary correspondence even for non-conserved topological spin currents. We note that a prior attempt was made to carefully examine the validity of bulk-boundary correspondence using spin pumping measurements in TI/insulating ferromagnet (Bi$_{1-x}$Sb$_x$Te/YIG) heterostructures ~\cite{HWang_PhysRevResearch.1.012014}. However, obtaining a reasonable agreement between theory and experiment in that study required incompletely validated assumptions about the location of the chemical potential in the experiments.  

Apart from testing bulk-boundary correspondence, the results presented here are also of contemporary translational interest for energy efficient non-volatile magnetic random access memory~\cite{Hellman_RevModPhys.89.025006,Miron_RevModPhys.91.035004,PhysRevLett.106.036601LiuSTFMR,Miron_NatMat,Liu2012}. Topological materials continue to play an important role in this context via ``topological spintronics'' because the bulk-boundary correspondence implies the presence of Berry curvature in the entire band structure lends itself naturally to generating strong SOTs. To date, large SOTs have been reported in TIs~\cite{Mellnik2014,Kondou2016,DC2018,PhysRevLett.123.207205} and topological semimetals ~\cite{PhysRevApplied.16.054031yanez,Yanez_PhysRevApplied.18.054004}. Efficient current-induced magnetization switching has also been demonstrated in TI/ferromagnet heterostructures at cryogenic temperatures~\cite{Fan_NMat_2014} and at room temperature~\cite{Han_PhysRevLett.119.077702,DC2018}. An accurate measurement of charge-to-spin conversion in \BiSb,   
the first discovered three-dimensional TI, is thus important.

We grow the thin films of \BiSb~on (001) sapphire substrates by co-deposition from Bi and Sb effusion cell sources in a Scienta-Omicron EVO-50 MBE chamber with a base pressure of $\sim 1 \times 10^{-10}$ mbar. \BiSb~has a rhombohedral crystal structure belonging to the R3m space group. To enable good epitaxial growth of the \BiSb~layer, we first grow a thin (Bi$_{1-x}$Sb$_x)_2$Te$_3$ (BST) buffer layer (thickness $\sim 3$ quintuple layers, QLs) on sapphire before the growth of \BiSb~(Fig. 1(a)). 

We note that BST itself is a TI, but at the 3 QL thickness needed for full epitaxial coverage, it is insulating even at 300 K. This allows us to grow \BiSb~layers on top without a current shunting effect for our experiments. We monitor the MBE growth using 13 keV reflection high energy diffraction (RHEED). Figure 1(b) shows the sharp streaky patterns of both the BST layer and the \BiSb~layer grown on it, indicating the epitaxial growth of both layers. A capping layer of about 40 nm tellurium is grown on top for {\it ex situ} characterization. We calibrate the growth rate of the materials by atomic force microscopy and determine sample composition using x-ray photoemission spectroscopy (XPS). Extensive {\it ex situ} characterization via x-ray diffraction (XRD) provides additional information about the structural characteristics of our \BiSb~thin films. Figure 1(c) shows an example of the XRD pole figures measured on one sapphire/BST/Bi$_{0.84}$Sb$_{0.16}$ (10nm)/Te sample. The presence of the very thin BST cannot be detected in its signal in the pole figures. However, these plots clearly visualize the signals from both the sapphire substrate and the Bi$_{0.84}$Sb$_{0.16}$ layer. All the signal peaks from Bi$_{0.84}$Sb$_{0.16}$ are well-focused, indicating a high crystalline quality of the thin film. As shown in Fig. 1(c), three of the Bi$_{0.84}$Sb$_{0.16}$ peaks are on top of the sapphire peaks, while the other three Bi$_{0.84}$Sb$_{0.16}$ peaks are rotated by $60^{\circ}$, suggesting some degree of in-plane twinning growth in the \BiSb~thin film. More detailed information about sample characterization using XRD and Raman spectroscopy is reported elsewhere \cite{Huang_2023}. For SOT measurements, we deposit a layer of Ni$_{0.8}$Fe$_{0.2}$ (permalloy, Py) on top of the \BiSb~and a zirconium capping layer, both {\it in situ} using e-beam evaporation. Finally, we characterize the BST/\BiSb/Py heterostructures using aberration-corrected scanning transmission electron microscopy (STEM). Figure 1(d) shows the cross-sectional high-angle annular dark-field (HAADF) STEM image of a BST/Bi$_{0.84}$Sb$_{0.16}$/Py sample, which shows the crystal structure of the epitaxial BST and \BiSb~layers and the amorphous Py. The atomically flat interface between the BST and \BiSb~layers again evidences the good epitaxial growth of \BiSb. 

\begin{figure}
    \includegraphics[width=15cm]{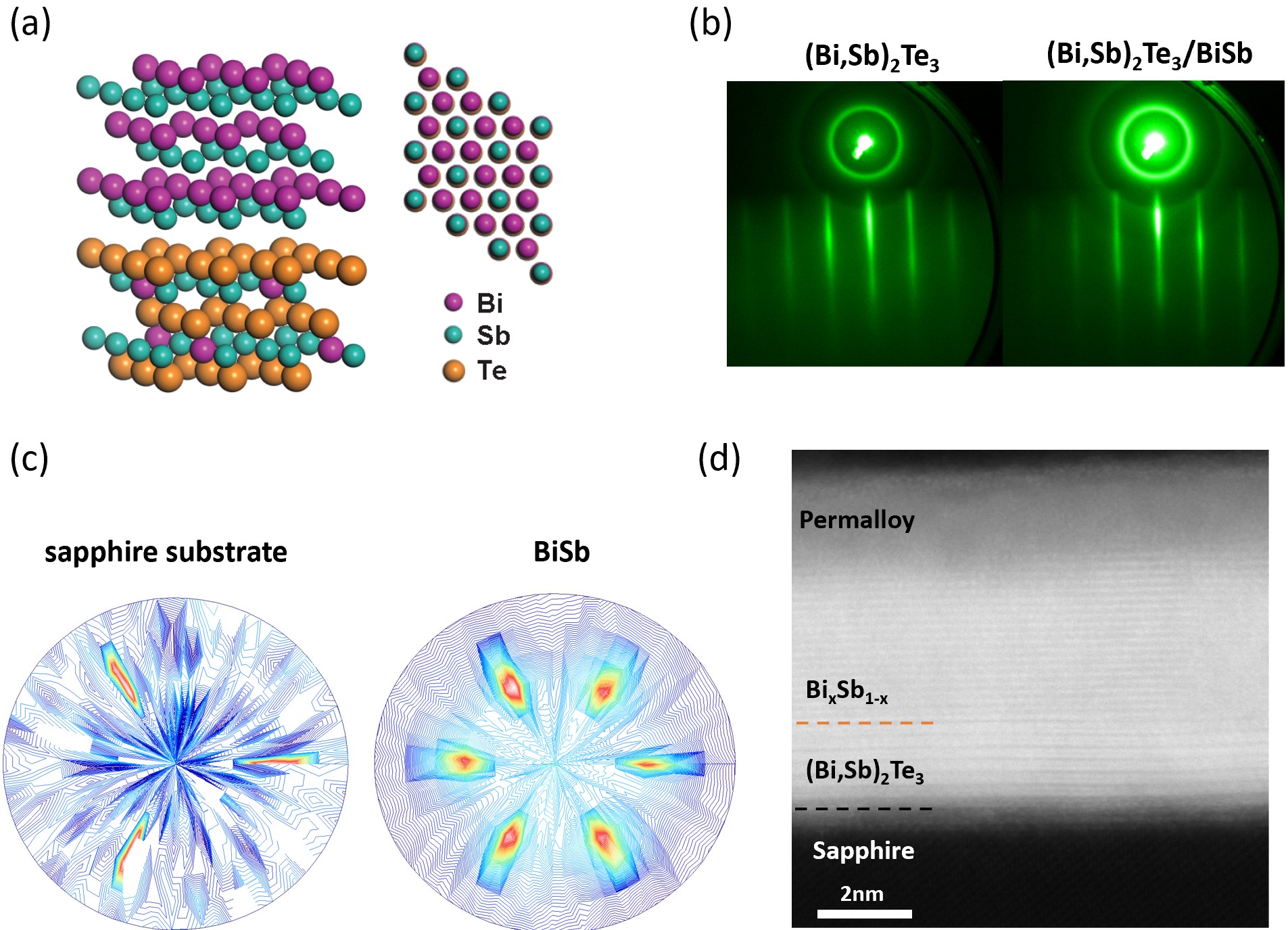}
    \caption{\textbf{Structural characterisation} (a) Schematic of BST/\BiSb~heterostructure. (b) RHEED patterns of the BST and the Bi$_{0.84}$Sb$_{0.16}$ layers. (c) XRD pole figures of the sapphire substrates and the 10 nm Bi$_{0.84}$Sb$_{0.16}$ thin film. (d) HAADF STEM image of the BST/Bi$_{0.84}$Sb$_{0.16}$/Py heterostructure in cross section. Scale bar is 2 nm.} 
    \label{1}
\end{figure}

We use \textit{in vacuo} ARPES to characterize the band structure of the various \BiSb~thin films grown in this study. Figure 2(a) shows the projected rhombohedral Brillouin zone along the \BiSb~[111] direction as well as the ARPES mapping of the Fermi surface of a Bi$_{0.84}$Sb$_{0.16}$ sample. Because our \BiSb~samples are deposited on a BST buffer layer, they also inherit the out-of-plane [111] crystalline orientation in the growth. The hexagonal six-fold symmetry in the Fermi mapping (Fig. 2(a)), with the extended band structure along the $\bar{\Gamma} - \bar{M}$ direction, is consistent with its out-of-plane direction indicated by the projected Brillouin zone and agrees well with previous ARPES measurements of cleaved \BiSb~(111) crystals~\cite{Hsieh2008} and much thicker \BiSb~films grown by MBE~\cite{Benia2015}. The band structure for a pure Bi sample, a Bi-rich Bi$_{0.84}$Sb$_{0.16}$ sample with the composition inside the TI regime, and a pure Sb sample are shown in Figs. 2(c) and (d) respectively. All three samples have a thickness of 10 nm. Figure 2(c) shows that the band structure of pure Bi and Bi$_{0.84}$Sb$_{0.16}$ is similar. However, the band structure of the pure Sb sample shows a clear change from that of pure Bi and Bi$_{0.84}$Sb$_{0.16}$ because the bulk valence band at the $\bar{\Gamma}$ point shifts more substantially toward the lower binding energy for pure Sb~\cite{Benia2015}. This renders a different behavior of the positions and shapes of the surface states near the $\bar{\Gamma}$ point for Bi-rich versus Sb-rich compositions. We also measured a control sample of only the BST (~3u.c.) layer in Fig. 2(b), which shows a drastically different ARPES spectrum compared to the \BiSb~ones in Fig. 2(c). In early ARPES measurements of bulk-cleaved \BiSb~crystals, three spin-polarized surface states were identified as the topological surface states in \BiSb~\cite{Hsieh2008}; one of these was later attributed to surface defects\cite{Benia2015,Zhang_PhysRevB.80.085307}. To better visualize the topological surface states in our epitaxial \BiSb~thin films, we plot the ARPES spectra second-derivatives with respect to the energy in Fig. 2(d). This figure shows that, for Bi-rich samples, the crossing of the two surface states at the $\bar{\Gamma}$ point cannot be discerned due to the spectral weight loss known in the pure Bi(111) surface band structure\cite{Benia2015,Ast_PhysRevB.67.113102}. Nevertheless, the two surface states in the Bi-rich samples can still be recognized in the vicinity of the $\bar{\Gamma}$ point. For the pure Sb sample, both the two surface states and their crossing at the $\bar{\Gamma}$ point are more clearly visualized. These ARPES measurements of the band structure, particularly the two surface states, in our \BiSb~thin films are in good agreement with previous studies on bulk-cleaved crystals~\cite{Hsieh2008} and MBE-grown thin films~\BiSb~\cite{Benia2015}, consistent with the reasonable crystallinity in our samples. This provides a firm foundation for us to compare the SOT measurements in these \BiSb~thin films to theoretical predictions based on a pristine \BiSb~crystal structure. The only cautionary remark is that ARPES measurements cannot provide any information about the influence of the ferromagnetic overlayer on the Dirac surface states at the buried interface. Indeed, first principles calculations indicate that the helical Dirac surface states of a TI are strongly perturbed when interfaced with a metallic ferromagnet ~\cite{zhang2016band}.
\textbf{\begin{figure}
    \includegraphics[width=12cm]{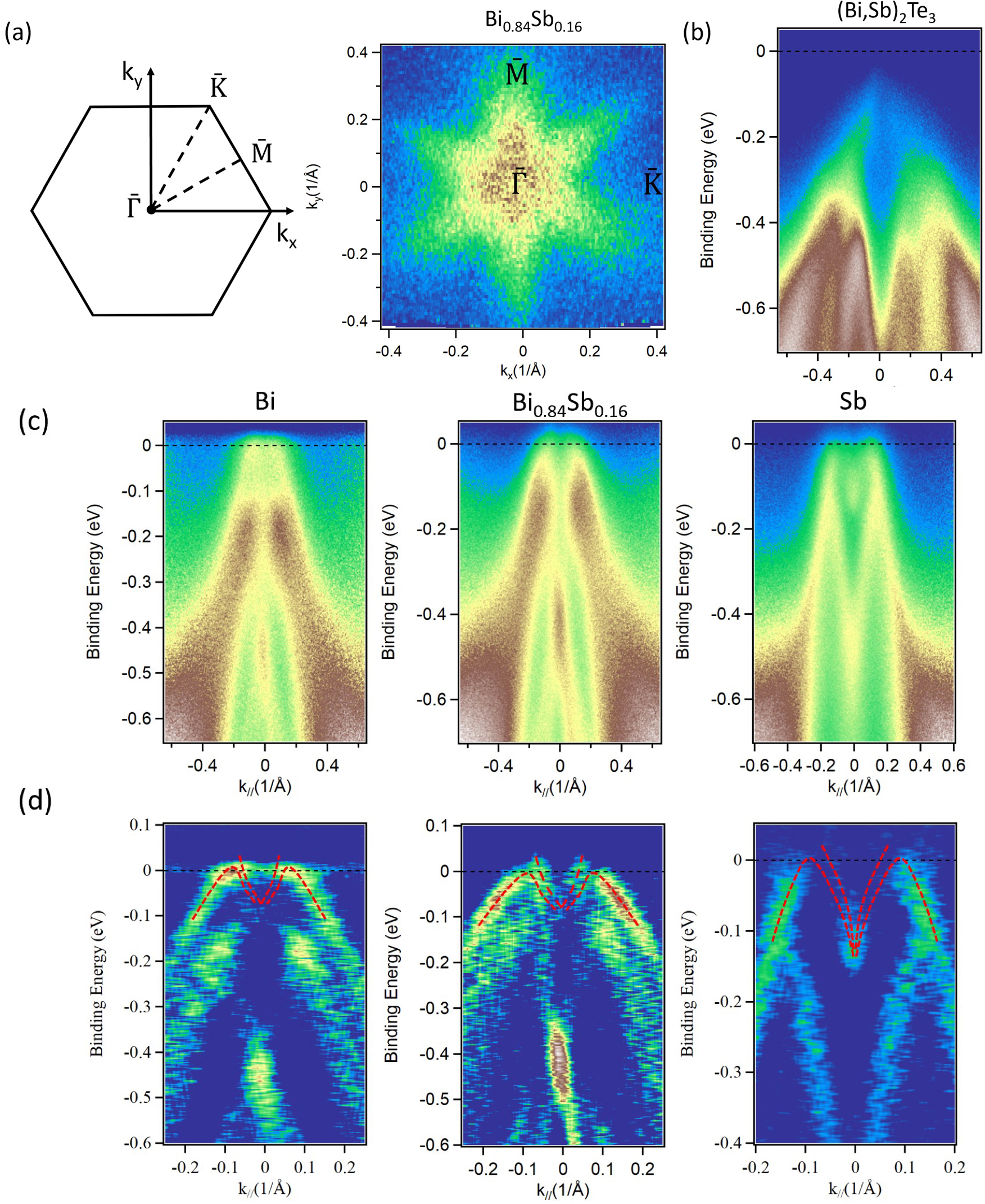}
    \caption{\textbf{ARPES characterization} (a) Projected Brillouin zone at the \BiSb(111) surface and the Fermi mapping of a Bi$_{0.84}$Sb$_{0.16}$(111) sample. (b) Band dispersion of a BST sample. (c) Band dispersion and (d) second-derivative plots of the ARPES spectra along the $\bar{K} - \bar{\Gamma} - \bar{K}$ direction for a Bi, a Bi$_{0.84}$Sb$_{0.16}$, and a Sb sample respectively. All the ARPES data were taken at 300 K with 21.2 eV photon excitation from a He lamp. The red dashed lines are guides to the eyes for the surface states in \BiSb.} 
    \label{2}
\end{figure}}

\textbf{\begin{figure}
    \includegraphics[width=12cm]{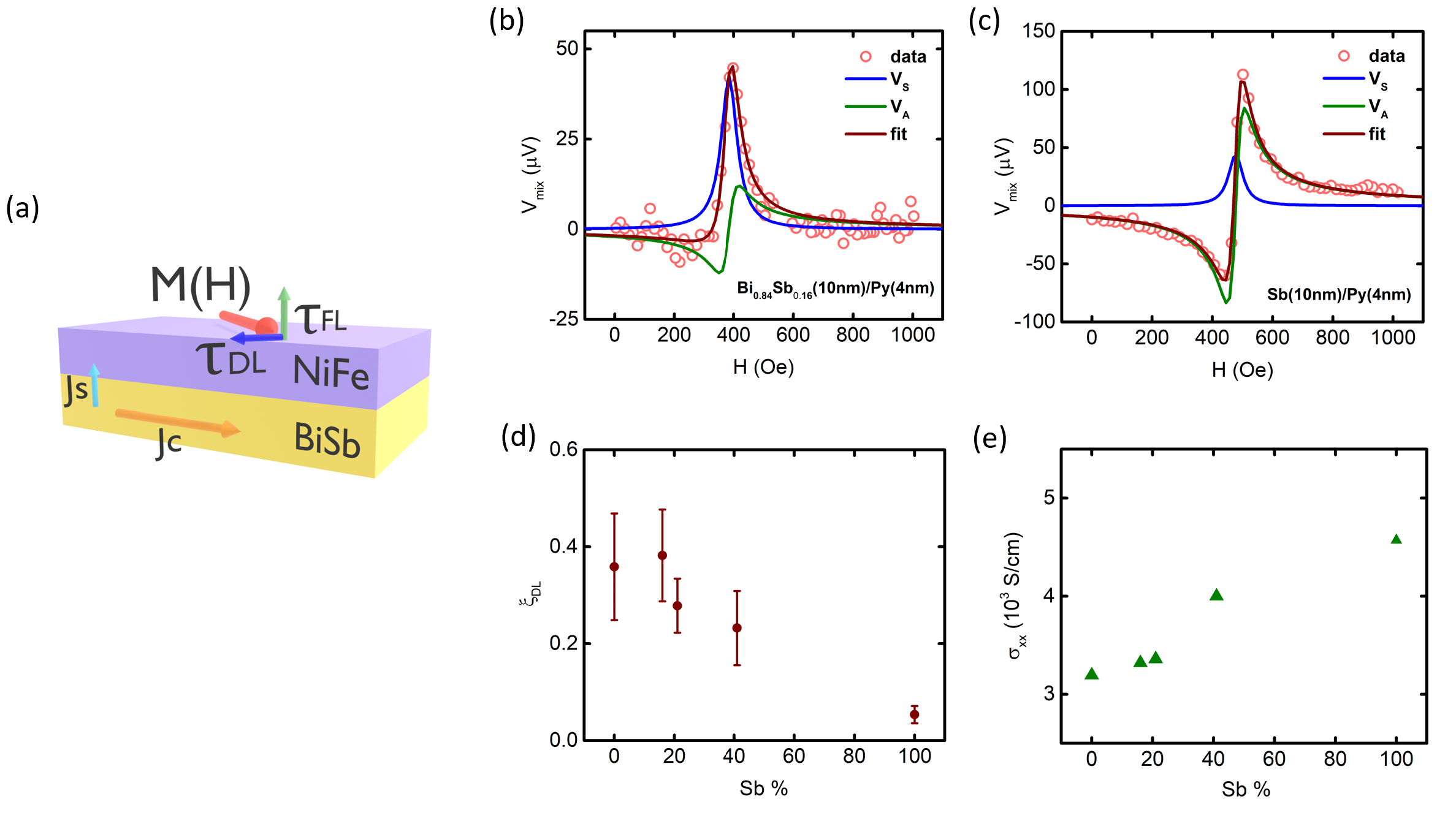}
    \caption{\textbf{ST-FMR measurements} (a) Schematic of the \BiSb/Py ST-FMR device. (b), (c) ST-FMR spectra measured at 6 GHz in a Bi$_{0.84}$Sb$_{0.16}$(10nm)/Py(4nm) and a Sb(10nm)/Py(4nm) sample, respectively. (d),(e) Dampinglike spin torque efficiency $\xi_{DL}$ and longitudinal electrical conductivity $\sigma_{xx}$, respectively, as a function of \BiSb~alloy composition.} 
    \label{3}
\end{figure}}

We next focus on the SOT measurements on the \BiSb/Py heterostructures via ST-FMR. Figure 3(a) shows the schematic of an ST-FMR device with a lateral dimension of $50 \mu m \times 10 \mu m$ for the effective area. An rf frequency charge current, $J_c$, is sent through the heterostructure, while the rectified time-average voltage across the device, $V_{mix}$, is measured. In a material system with a SOT effect (in this case, the \BiSb~layer), the charge current induces a diffusive transverse spin current, $J_s$, which is injected into the adjacent FM layer, exerting a SOT on the magnetization. In ST-FMR, the strength of such a SOT effect can then be estimated via the spin torque efficiency, $\xi_{FMR}$, given by~ \cite{PhysRevLett.106.036601LiuSTFMR,Pai_PhysRevB.92.064426}:

\begin{equation}
   \xi_{\mathrm{FMR}}= \frac{V_S}{V_A} \left(\frac{e}{\hbar}\right) \mu_0 M_S t_{\mathrm{BiSb}}t_{\mathrm{F}} \left[1+\left(\frac{M_{\mathrm{eff}}}{H_{\mathrm{res}}}\right)\right]^{1/2}.
   \label{eq:xi}
\end{equation}               

Here, $e$ is the charge of the electron, $\hbar$ is the reduced Planck constant, $\mu_{0}$ is the permeability of free space, $M_S$ is the saturation magnetization of Py, $t_\mathrm{{(BiSb)(F)}}$ is the thickness of the \BiSb~(Py) layer, $M_{eff}$ is the effective magnetization and $H_{res}$ is the resonance field, respectively. $V_{S(A)}$ corresponds to the symmetric (antisymmetric) component of the line shape of $V_{mix}$ at the ferromagnetic resonance. The symmetric signal ($V_S$) results from the dampinglike spin torque, i.e., the spin torque exerted on the FM after the transverse component of the spin current is absorbed. The antisymmetric signal ($V_A$) comes from the field-like torque generated by the sum of any current-induced field components that include the Oersted field and other possible effective fields~\cite{Ou_PhysRevB.94.140414}. In our experiments, we prepared five sets of \BiSb/Py samples consisting of the following compositions: pure Bi, Bi$_{0.84}$Sb$_{0.16}$, Bi$_{0.79}$Sb$_{0.21}$, Bi$_{0.59}$Sb$_{0.41}$, and pure Sb, covering the whole TI and non-TI regimes of \BiSb. We note that two compositions, Bi$_{0.84}$Sb$_{0.16}$ and Bi$_{0.79}$Sb$_{0.21}$, are within the TI regime. Each set (composition) consists of three samples of 3 nm, 4 nm, and 5 nm Py respectively, while we fix all the \BiSb~layers’ thickness to be 10 nm. 

Figures 3(b) and (c) show the ST-FMR spectra of a Bi$_{0.84}$Sb$_{0.16}$(10nm)/Py(4nm) sample and an Sb(10nm)/Py(4nm). For the Bi$_{0.84}$Sb$_{0.16}$~ sample, the fit to the resonance spectra reveals a larger symmetric component $V_S$ originating from the dampinglike torque than the antisymmetric component $V_A$ from the field-like torque. In comparison, for the pure Sb sample, its $V_S$(dampinglike) component is comparatively smaller than the $V_A$(field-like) component, suggesting that the strength of the dampinglike SOT in Bi-rich Bi$_{0.84}$Sb$_{0.16}$~ is stronger than pure Sb. It is known that in ST-FMR measurements, if the effective field component(s) other than the Oersted field cannot be ignored, the computed $\xi_{FMR}$ from Eq.(1) usually depends on the thicknesses of the FM, $t_F$~\cite{Yanez_PhysRevApplied.18.054004,Pai_PhysRevB.92.064426,An2016CuOx}. In the five sets of samples we measured, for each composition of \BiSb, we did not see a Py thickness dependence of $\xi_{FMR}$, indicating the only field-like component in $V_A$ is from the Oersted field~\cite{Pai_PhysRevB.92.064426}. The $t_F$ independence also makes $\xi_{FMR}$ a direct measurement of the dampinglike spin torque efficiency, i.e., $\xi_{FMR}=\xi_{DL}$.

We summarize the dampinglike spin torque efficiency $\xi_{DL}$ of each \BiSb~composition in Fig. 3(d). In the Bi-rich region, the three compositions, pure Bi, Bi$_{0.84}$Sb$_{0.16}$, and Bi$_{0.79}$Sb$_{0.21}$, all show comparable $\xi_{DL}$, with their strength ranging between around 0.3 and 0.4. These dampinglike spin torque efficiencies are stronger than, or at least comparable to, those in heavy metals ~\cite{PhysRevLett.106.036601LiuSTFMR,Pai2012} but noticeably smaller than the spin Hall angles or SOT efficiencies previously reported in some epitaxial or non-epitaxial \BiSb~\cite{Khang2018,Khang2020,Fan2022}. As shown in Fig. 3(d), as the Sb concentration further increases, $\xi_{DL}$ decreases in a quasi-linear fashion towards its minimum at pure Sb. While the overall trend of this composition-dependent dampinglike spin torque efficiency is consistent with the relative strength of the spin-orbit interaction in Bi and Sb, at first glance, its magnitude appears to peak at the Bi$_{0.84}$Sb$_{0.16}$~ composition, making one wonder whether the TI state plays a role in generating the SOTs. To resolve this, we use the ST-FMR data to extract the effective spin Hall conductivity, $\sigma_{SH}^{eff} = \frac{\hbar}{2e} \xi_{DL} \sigma_{xx}$, and compare the experimental results with theoretical calculations. We note that $\sigma_{SH}^{eff}$ is a better parameter to evaluate the physical origin of the SOTs since $\xi_{DL}$ itself depends on the electrical conductivity of the SOT material. 

Figure 4 (a) shows the theoretically calculated SHC as a function of \BiSb~composition for the crystalline orientation and measurement geometry of the films in our experiments ($\sigma ^{z}_{yx}$, green curve); we also show the calculated SHC for two orthogonal geometries (see Appendix A for details). Note that this calculation only uses the bulk states of \BiSb. To compute $\sigma_{SH}^{eff}$ from experimental data, we first determine the electrical conductivity ($\sigma_{xx}$) for each \BiSb~composition. We grow the \BiSb~(10nm) samples with ~40 nm tellurium as a capping layer and perform the transport measurements in a PPMS system. Figure 3(e) shows $\sigma_{xx}$ as a function of \BiSb~composition. We then compute the effective SHC $\sigma_{SH}^{eff}$  based on the spin-torque and transport measurements, with the results shown in Fig. 4(b). The overall composition dependence of $\sigma_{SH}^{eff}$ , particularly in the regime with Sb concentration less than ~40 \%, is more gradual than the dependence in $\xi_{DL}$ (Fig. 3(d)), as the smaller $\xi_{DL}$ in higher Sb concentration samples are compensated by the larger electrical conductivity in the computation of $\sigma_{SH}^{eff}$. Figure 4(b) also directly compares the experimental results with the theoretically predicted variation of the SHC ($\sigma ^{z}_{yx}$) of \BiSb~with alloy composition, showing excellent agreement. This provides strong evidence supporting the intrinsic spin Hall effect as the physical origin of the observed SOT effect in epitaxial \BiSb~thin films. The results here also indicate that the SOT effect in epitaxial \BiSb~is dominated by the strong spin-orbit entanglement of bulk states well below the chemical potential: this connects directly to the SHC in epitaxial \BiSb~films and is responsible for generating the SOT exerted on a vicinal metallic ferromagnet. In a fundamental sense, bulk-boundary correspondence makes it moot to ask whether the enhanced SHC and spin-charge conversion efficiency observed in many topological quantum materials originates in the bulk states or in the surface states.     

\textbf{\begin{figure}
    \includegraphics[width=6cm]{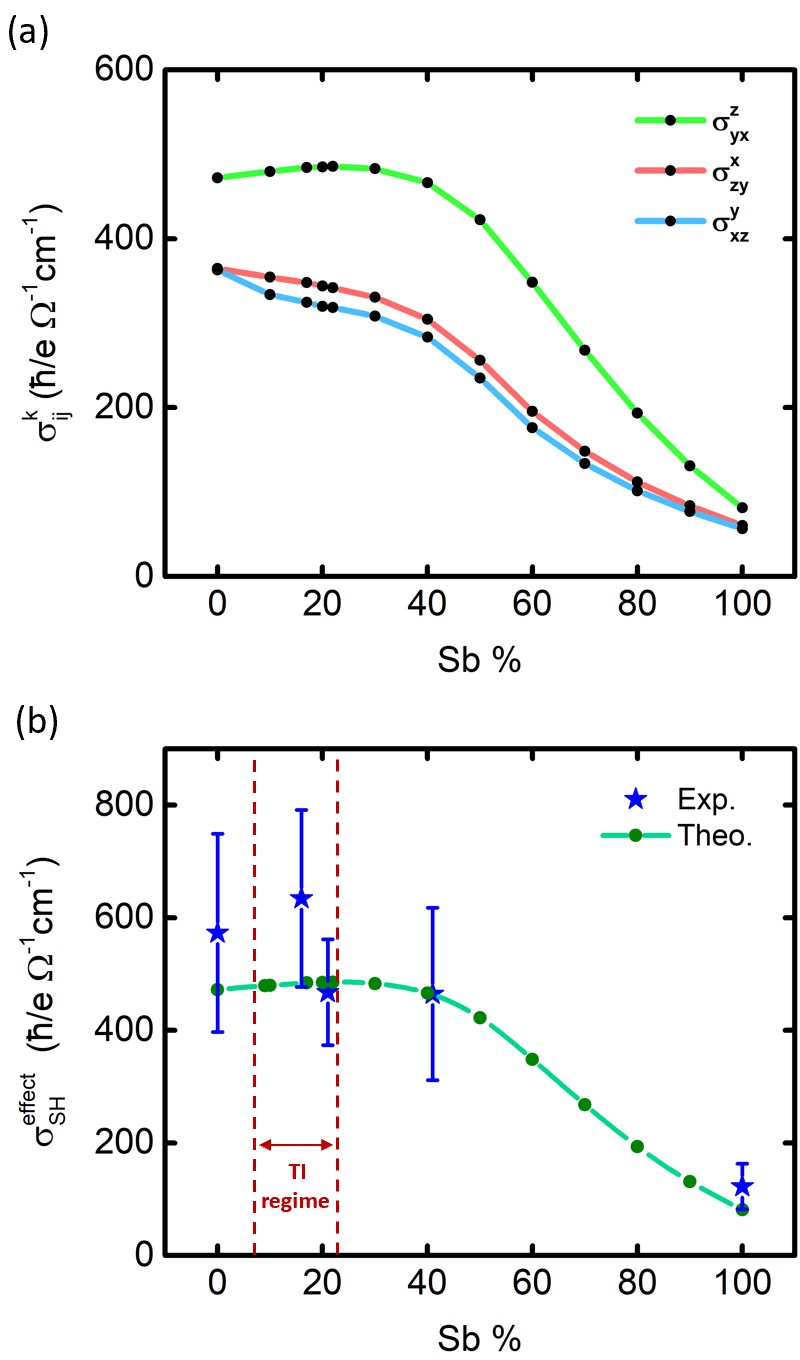}
    \caption { \textbf{Spin Hall conductivity: theory vs. experiment} (a) SHC calculated from the tight-binding Hamiltonian for bulk states in \BiSb~as a function of composition. (b) Comparison between experimentally determined effective spin Hall conductivity $\sigma_{\mathrm{SH}}^{\mathrm{eff}}$ (star symbol) in \BiSb~as a function of composition (as determined from ST-FMR and the calculated spin Hall conductivity $\sigma^{z}_{yx}$ (from panel (a)). The composition range corresponding to the TI regime is highlighted. } 
    \label{3}
\end{figure}}
Now we address the discrepancy in the strength of SOTs observed in \BiSb~in previous studies \cite{Khang2018,Chi2020}. The SHC reported in MBE-grown \BiSb/MnGa heterostructures in Ref.~\cite{Khang2018} is at least two orders of magnitude larger than the computed values in our epitaxial \BiSb. Since the crystalline orientation of the Bi$_{0.9}$Sb$_{0.1}$ surface in Ref.~\cite{Khang2018} is [012] (rather than [111] as in our case), one might hypothesize that this surface hosts more Dirac cones than other orientations. However, we tested this hypothesis by calculating the SHC for this orientation (data not shown) and found that it is only enhanced by $\sim 20\%$.  We note that the SOT efficiency estimated in Ref.~\cite{Khang2018} relies on the current-assisted magnetic field switching assuming a macro-spin model; this may result in larger uncertainty than expected because it is essentially a SOT-assisted magnetization switching measurement. In such switching experiments, the ferromagnet is usually in a multidomain magnetic structure instead of a simple macro spin state~\cite{Lee_PhysRevB.89.024418}. Therefore, the multidomain switching process, typically involving domain wall motion, may complicate the analysis in estimating the SOT efficiency. Finally, the role of thermally assisted switching is difficult to definitively rule out and the resulting error in overestimated switching efficiency is hard to quantify. This has been noted in another study that used the harmonic response technique to report no SOT effect in Bi$_{0.74}$Sb$_{0.26}$/Co heterostructures~\cite{Roschewsky2019}. Instead, this study showed a significant spurious ordinary Nernst effect dominating the measured signals, leading to the overestimation of the spin Hall angle and SHC. In that work, however, the Bi$_{0.74}$Sb$_{0.26}$ layer was grown by MBE and the ferromagnetic Co layer was deposited {\it ex situ}. The effect of the air-exposed surface of \BiSb~could affect the measurement of the actual SOTs.

It is difficult to directly compare our results to the previous report based on sputtered polycrystalline \BiSb~thin films~\cite{Chi2020} because the band structure in those samples is unknown. The discrepancies between that study and ours are more dificult to reconcile due to the possibly different sample conditions and experimental settings. For example, Ref.~\cite{Chi2020} reports strong temperature-dependent SOTs in \BiSb/CoFeB heterostructures. The authors tentatively attribute the temperature-dependent SHC behavior in \BiSb~to the thermally excited Dirac-like $L$ electrons in that work. But other extrinsic origins to the SHC, such as the side-jump mechanism, cannot be excluded. Measurements of ST-FMR in our samples do not show any appreciable temperature dependence of the SHC (Appendix B). It is also unknown how quantum confinement due to the reduced crystalline domain sizes in sputtered \BiSb~samples may affect the SOT behavior, as reported in sputtered Bi$_2$Se$_3$\cite{DC2018}. We do note that the harmonic response technique adopted by the prior studies of SOT efficiency in sputtered \BiSb~suffer the same concerns about spin Nernst contributions pointed out earlier ~\cite{Roschewsky2019}.

In conclusion, we synthesize epitaxial \BiSb~thin films using MBE. We measure the SOTs in \BiSb~/Py heterostructures prepared {\it in situ} and compare our results to first-principle calculations. We confirm the topological surface states in our epitaxial \BiSb~films via {\it in vacuo} ARPES. By varying the thickness of Py, we determine the dampinglike spin torque efficiencies for different \BiSb~compositions from pure Bi to pure Sb. We observe relatively strong dampinglike SOTs in Bi-rich \BiSb~alloys that are comparable to heavy metals. However, compared to some earlier reports, we do not observe a particularly gigantic SOT efficiency or SHC in our epitaxial \BiSb. Instead, the estimated SHC results in our epitaxial \BiSb~are in excellent agreement with tight binding calculations of the intrinsic SHC from bulk states in \BiSb, thus strongly supporting bulk-boundary correspondence. Our work is a systematic study investigating the SOTs in TI material \BiSb~and calls for caution in studying the SOT effects in other TI and topological systems.

\begin{acknowledgments}
The authors would like to thank A. Sengupta for providing access to apparatus used in ST-FMR measurements. The principal support for this project was provided by the Penn State Two-Dimensional Crystal Consortium-Materials Innovation Platform (2DCC-MIP) under NSF Grant No. DMR-2039351 (YO, AR, NS). Additional support was provided by SMART, one of seven centers of nCORE, a Semiconductor Research Corporation program, sponsored by the National Institute of Standards and Technology (NIST) (WY, NS, SG, AM) and NSF Grant No. DMR-2309431 (SG, KAM). WY also acknowledges partial support from the Penn State MRSEC (DMR-2011839) for ST-FMR measurements performed during the completion of this paper. Parts of this work were carried out in the Characterization Facility, University of Minnesota, which receives partial support from the NSF through the MRSEC (DMR-2011401).
\end{acknowledgments}


\appendix

\section{Theory}
We calculate the intrinsic SHC using the Kubo formula in the linear response theory regime and clean static limit. The intrinsic SHC is the sum over the momentum resolved Berry curvatures, $\Omega_n^k (\v k)$, of all filled bands such as:
\al{\label{eq:shc}
\sigma_{ji}^k=\frac{e\hbar}{V}\sum_{\v k}\sum_n f_{\v k n}\Omega_n^k (\v k),}
and
\al{\label{eq:curvature}
\Omega_n^k (\v k) =2\sum_{n \neq n'} {\rm Im} \frac{\big<{u_{n \v k}}|{\hat j_j^k}| {u_{n' \v k}}\big>\big< {u_{n' \v k}}|{\hat v_i}|{u_{n \v k}}\big>}{(E_{n\v k} -E_{n' \v k})^2}.
}
Here, $i$,$j$, and $k$ correspond to the $\hat{x}, \hat{y}$, and $\hat{z}$ crystallographic directions, $V$ is the volume of the system, $\hat v_i$ is the velocity operator along the $i$ direction and is linked to the derivative of the Hamiltonian with respect to $k_i$ ($\hbar \hat v_i=\nabla_{k_i}\hat H$), and $\hat j_i^j=\frac{\hbar}{4}(\hat v_i\sigma_j+\sigma_j\hat v_i)$ is the spin current operator.  The eigenvalues ($E_{n\v k}$) and wave functions ($u_{n\v k}$) are extracted from the construction of a tight-binding Hamiltonian including spin-orbit coupling parameters\cite{LiuAllen}.
In order to calculate the SHC of the Bi$_{1-x}$Sb$_x$ alloy, we use a linear virtual crystal approximation and mix the tight-binding parameters of Bi and Sb according to the alloy composition. As shown in Fig. 4(a), the largest SHC is for the $yxz$ configuration for all compositions, which corresponds to a charge current along the $x$ direction, a spin current in the $y$ direction, and spin polarization along the $z$ direction. The other two directions also exhibit significantly large SHC with minimum anisotropy. In general, Sb exhibits lower anisotropy, and the value of SHC drops monotonically as alloyed with Sb following decreasing effective spin-orbit coupling with the alloy composition. For the $yxz$ configuration, the largest SHC can be observed between 10-22 \% of antimony, which also corresponds to the semiconductor phase and observed topological insulator phase\cite{Hsieh2008}. The largest value of the SHC we obtained is around 480 $(\hbar/e)(\Omega^{-1}\text{cm}^{-1})$. 

\section{Temperature dependence of ST-FMR}

Temperature-dependent ($290$~K to $3.6$~K) ST-FMR measurements are performed in a closed cycle Montana cryostat. The sample is placed on a custom-designed designed holder, and the excitation frequency is kept fixed at $f=3$~GHz during the measurement cycle. The magnitude of $V_{mix}$ obtained from a Bi$_{0.84}$Sb$_{0.16}$(10 nm)/Py(4 nm) sample measured films does not show any apparent temperature dependence of the raw signal (Fig. 5). Measurements carried out on another similar Bi$_{0.84}$Sb$_{0.16}$~sample but with a 5 nm Py overlayer show similar behavior. 

\begin{figure}[h!]
   \centering
    \includegraphics[width=14cm]{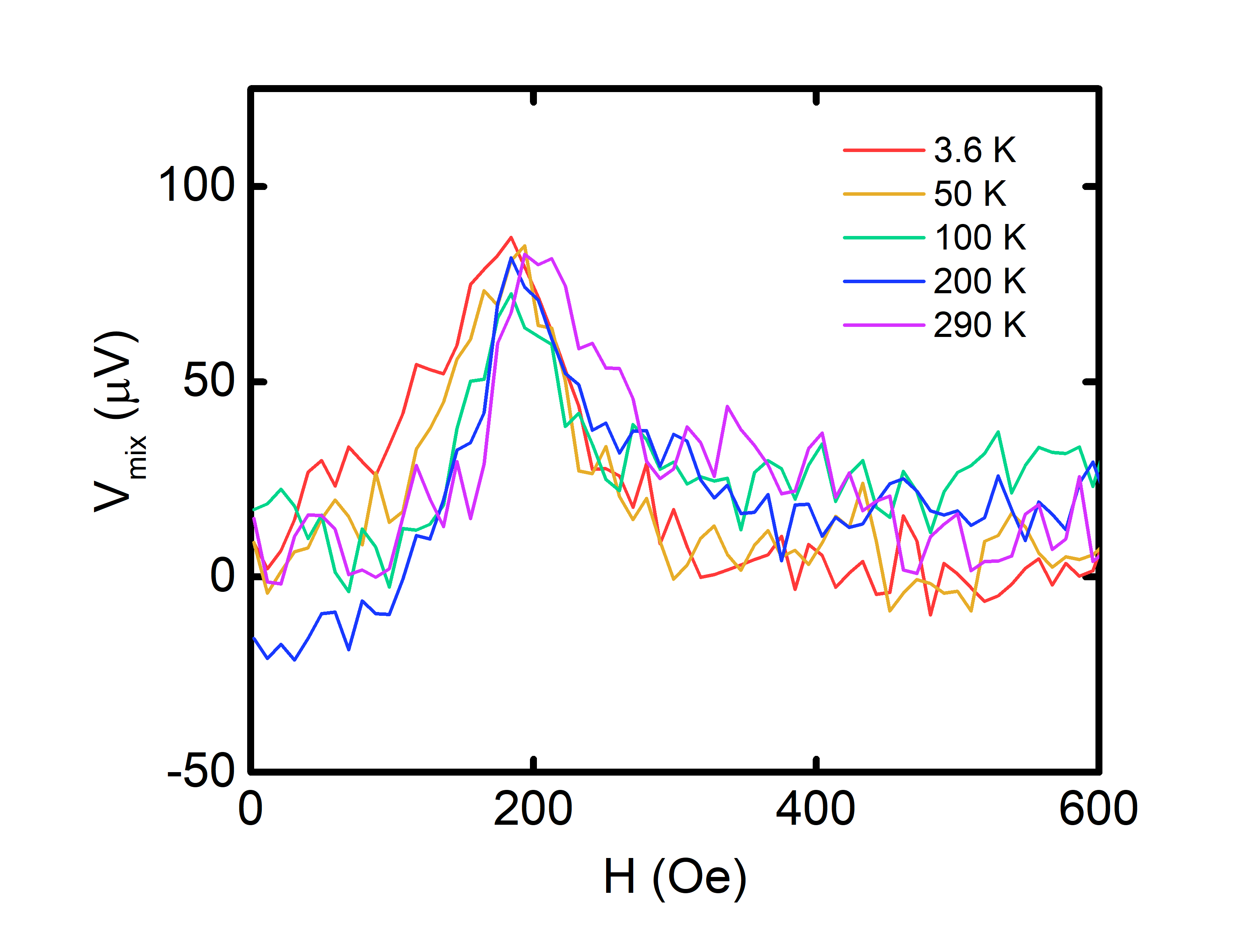}
    \caption{$V_{mix}$ at different temperatures for a Bi$_{0.84}$Sb$_{0.16}$(10 nm)/Py(4 nm) sample.}
    \label{SI1}
\end{figure}

\newpage
\section{Summary of samples}

Table I summarizes the MBE growth conditions for all the ST-FMR samples presented in this work. The data associated with these samples can be publicly accessed through the 2D Crystal Consortium’s Lifetime Sample Tracking (LiST) data management system (https://m4-2dcc.vmhost.psu.edu/list/data/TQ2fnLIO1y1F).

\begin{table}

\begin{tabular}{|c|c|c|c|c|c|}
\hline 
Sample ID & Structure & Bi cell & Sb cell & \BiSb~growth\tabularnewline
&  & temperature $(^{\circ} \mathrm{C})$ & temperature $(^{\circ} \mathrm{C})$ & temperature $(^{\circ} \mathrm{C})$\tabularnewline
\hline 
\hline 
220407A-YO &	Bi(10nm)/Py(3nm) &	470	& N/A &	90\tabularnewline
\hline 
220409A-YO &	Bi(10nm)/Py(4nm) &	470 &	N/A	& 90\tabularnewline
\hline 
220410B-YO	& Bi(10nm)/Py(5nm)	& 470 &	N/A	& 90\tabularnewline
\hline 
220508A-YO	& Bi$_{0.84}$Sb$_{0.16}$(10nm)/Py(3nm) &	470 &	316 & 160\tabularnewline
\hline
220509B-YO &	Bi$_{0.84}$Sb$_{0.16}$(10nm)/Py(4nm) &	470 &	316 &	160\tabularnewline
\hline
220508B-YO &	Bi$_{0.84}$Sb$_{0.16}$(10nm)/Py(5nm)	& 470 &	316 &	160\tabularnewline
\hline
220428B-YO	& Bi$_{0.79}$Sb$_{0.21}$(10nm)/Py(3nm)	& 470 &	320 &	160\tabularnewline
\hline
220501B-YO &	Bi$_{0.79}$Sb$_{0.21}$(10nm)/Py(4nm)	& 470 &	320 &	160\tabularnewline
\hline
220501C-YO	& Bi$_{0.79}$Sb$_{0.21}$(10nm)/Py(5nm)	& 470 &	320 &	160\tabularnewline
\hline
220512B-YO	& Bi$_{0.59}$Sb$_{0.41}$(10nm)/Py(3nm) &	470 &	336 &	160 \tabularnewline
\hline
220516A-YO &	Bi$_{0.59}$Sb$_{0.41}$(10nm)/Py(4nm)	& 470 &	336 &	160\tabularnewline
\hline
220518A-YO &	Bi$_{0.59}$Sb$_{0.41}$(10nm)/Py(5nm) &	470 &	336 &	160 \tabularnewline
\hline
220410A-YO	& Sb(10nm)/Py(3nm) &	N/A &	337 &	160\tabularnewline
\hline
220425A-YO	& Sb(10nm)/Py(4nm) &	N/A	& 337 &	160\tabularnewline
\hline
220425B-YO & Sb(10nm)/Py(5nm) &	N/A	& 337	& 160\tabularnewline
\hline
\end{tabular}

\caption{MBE growth parameters for all the ST-FMR samples presented in this work. The growth temperatures were monitored using an Optris Xi80 IR camera.}
\end{table}

\newpage


\begin{thebibliography}{41}%
\makeatletter
\providecommand \@ifxundefined [1]{%
 \@ifx{#1\undefined}
}%
\providecommand \@ifnum [1]{%
 \ifnum #1\expandafter \@firstoftwo
 \else \expandafter \@secondoftwo
 \fi
}%
\providecommand \@ifx [1]{%
 \ifx #1\expandafter \@firstoftwo
 \else \expandafter \@secondoftwo
 \fi
}%
\providecommand \natexlab [1]{#1}%
\providecommand \enquote  [1]{``#1''}%
\providecommand \bibnamefont  [1]{#1}%
\providecommand \bibfnamefont [1]{#1}%
\providecommand \citenamefont [1]{#1}%
\providecommand \href@noop [0]{\@secondoftwo}%
\providecommand \href [0]{\begingroup \@sanitize@url \@href}%
\providecommand \@href[1]{\@@startlink{#1}\@@href}%
\providecommand \@@href[1]{\endgroup#1\@@endlink}%
\providecommand \@sanitize@url [0]{\catcode `\\12\catcode `\$12\catcode
  `\&12\catcode `\#12\catcode `\^12\catcode `\_12\catcode `\%12\relax}%
\providecommand \@@startlink[1]{}%
\providecommand \@@endlink[0]{}%
\providecommand \url  [0]{\begingroup\@sanitize@url \@url }%
\providecommand \@url [1]{\endgroup\@href {#1}{\urlprefix }}%
\providecommand \urlprefix  [0]{URL }%
\providecommand \Eprint [0]{\href }%
\providecommand \doibase [0]{https://doi.org/}%
\providecommand \selectlanguage [0]{\@gobble}%
\providecommand \bibinfo  [0]{\@secondoftwo}%
\providecommand \bibfield  [0]{\@secondoftwo}%
\providecommand \translation [1]{[#1]}%
\providecommand \BibitemOpen [0]{}%
\providecommand \bibitemStop [0]{}%
\providecommand \bibitemNoStop [0]{.\EOS\space}%
\providecommand \EOS [0]{\spacefactor3000\relax}%
\providecommand \BibitemShut  [1]{\csname bibitem#1\endcsname}%
\let\auto@bib@innerbib\@empty
\bibitem [{\citenamefont {Balasubramanian}\ \emph {et~al.}(1999)\citenamefont
  {Balasubramanian}, \citenamefont {Kraus},\ and\ \citenamefont
  {Lawrence}}]{VBala_WOS:000079938100080}%
  \BibitemOpen
  \bibfield  {author} {\bibinfo {author} {\bibfnamefont {V.}~\bibnamefont
  {Balasubramanian}}, \bibinfo {author} {\bibfnamefont {P.}~\bibnamefont
  {Kraus}},\ and\ \bibinfo {author} {\bibfnamefont {A.}~\bibnamefont
  {Lawrence}},\ }\bibfield  {title} {\bibinfo {title} {{Bulk versus boundary
  dynamics in anti-de Sitter spacetime}},\ }\href
  {https://doi.org/10.1103/PhysRevD.59.046003} {\bibfield  {journal} {\bibinfo
  {journal} {Phys. Rev. D}\ }\textbf {\bibinfo {volume} {59}},\ \bibinfo
  {pages} {046003} (\bibinfo {year} {1999})}\BibitemShut {NoStop}%
\bibitem [{\citenamefont {Hasan}\ and\ \citenamefont
  {Kane}(2010)}]{Hasan_RevModPhys.82.3045}%
  \BibitemOpen
  \bibfield  {author} {\bibinfo {author} {\bibfnamefont {M.~Z.}\ \bibnamefont
  {Hasan}}\ and\ \bibinfo {author} {\bibfnamefont {C.~L.}\ \bibnamefont
  {Kane}},\ }\bibfield  {title} {\bibinfo {title} {Colloquium: Topological
  insulators},\ }\href {https://doi.org/10.1103/RevModPhys.82.3045} {\bibfield
  {journal} {\bibinfo  {journal} {Rev. Mod. Phys.}\ }\textbf {\bibinfo {volume}
  {82}},\ \bibinfo {pages} {3045} (\bibinfo {year} {2010})}\BibitemShut
  {NoStop}%
\bibitem [{\citenamefont {Teo}\ and\ \citenamefont
  {Kane}(2010)}]{Teo_PRB_2010}%
  \BibitemOpen
  \bibfield  {author} {\bibinfo {author} {\bibfnamefont {J.~C.~Y.}\
  \bibnamefont {Teo}}\ and\ \bibinfo {author} {\bibfnamefont {C.~L.}\
  \bibnamefont {Kane}},\ }\bibfield  {title} {\bibinfo {title} {Topological
  defects and gapless modes in insulators and superconductors},\ }\href
  {https://link.aps.org/doi/10.1103/PhysRevB.82.115120} {\bibfield  {journal}
  {\bibinfo  {journal} {Phys. Rev. B}\ }\textbf {\bibinfo {volume} {82}},\
  \bibinfo {pages} {115120} (\bibinfo {year} {2010})}\BibitemShut {NoStop}%
\bibitem [{\citenamefont {Schindler}\ \emph {et~al.}(2018)\citenamefont
  {Schindler}, \citenamefont {Wang}, \citenamefont {Vergniory}, \citenamefont
  {Cook}, \citenamefont {Murani}, \citenamefont {Sengupta}, \citenamefont
  {Kasumov}, \citenamefont {Deblock}, \citenamefont {Jeon}, \citenamefont
  {Drozdov}, \citenamefont {Bouchiat}, \citenamefont {Gueron}, \citenamefont
  {Yazdani}, \citenamefont {Bernevig},\ and\ \citenamefont
  {Neupert}}]{Schindler_NatPhys_2018}%
  \BibitemOpen
  \bibfield  {author} {\bibinfo {author} {\bibfnamefont {F.}~\bibnamefont
  {Schindler}}, \bibinfo {author} {\bibfnamefont {Z.}~\bibnamefont {Wang}},
  \bibinfo {author} {\bibfnamefont {M.~G.}\ \bibnamefont {Vergniory}}, \bibinfo
  {author} {\bibfnamefont {A.~M.}\ \bibnamefont {Cook}}, \bibinfo {author}
  {\bibfnamefont {A.}~\bibnamefont {Murani}}, \bibinfo {author} {\bibfnamefont
  {S.}~\bibnamefont {Sengupta}}, \bibinfo {author} {\bibfnamefont {A.~Y.}\
  \bibnamefont {Kasumov}}, \bibinfo {author} {\bibfnamefont {R.}~\bibnamefont
  {Deblock}}, \bibinfo {author} {\bibfnamefont {S.}~\bibnamefont {Jeon}},
  \bibinfo {author} {\bibfnamefont {I.}~\bibnamefont {Drozdov}}, \bibinfo
  {author} {\bibfnamefont {H.}~\bibnamefont {Bouchiat}}, \bibinfo {author}
  {\bibfnamefont {S.}~\bibnamefont {Gueron}}, \bibinfo {author} {\bibfnamefont
  {A.}~\bibnamefont {Yazdani}}, \bibinfo {author} {\bibfnamefont {B.~A.}\
  \bibnamefont {Bernevig}},\ and\ \bibinfo {author} {\bibfnamefont
  {T.}~\bibnamefont {Neupert}},\ }\bibfield  {title} {\bibinfo {title}
  {Higher-order topology in bismuth},\ }\href
  {https://doi.org/10.1038/s41567-018-0224-7} {\bibfield  {journal} {\bibinfo
  {journal} {Nat. Phys.}\ }\textbf {\bibinfo {volume} {14}},\ \bibinfo {pages}
  {918} (\bibinfo {year} {2018})}\BibitemShut {NoStop}%
\bibitem [{\citenamefont {Weis}\ and\ \citenamefont
  {Von~Klitzing}(2011)}]{KvK}%
  \BibitemOpen
  \bibfield  {author} {\bibinfo {author} {\bibfnamefont {J.}~\bibnamefont
  {Weis}}\ and\ \bibinfo {author} {\bibfnamefont {K.}~\bibnamefont
  {Von~Klitzing}},\ }\bibfield  {title} {\bibinfo {title} {{Metrology and
  microscopic picture of the integer quantum Hall effect}},\ }\href
  {https://doi.org/10.1098/rsta.2011.0198} {\bibfield  {journal} {\bibinfo
  {journal} {Philos. Trans. A Math. Phys. Eng. Sci.}\ }\textbf {\bibinfo
  {volume} {369}},\ \bibinfo {pages} {3954} (\bibinfo {year}
  {2011})}\BibitemShut {NoStop}%
\bibitem [{\citenamefont {Uri}\ \emph {et~al.}(2020)\citenamefont {Uri},
  \citenamefont {Kim}, \citenamefont {Bagani}, \citenamefont {Lewandowski},
  \citenamefont {Grover}, \citenamefont {Auerbach}, \citenamefont {Lachman},
  \citenamefont {Myasoedov}, \citenamefont {Taniguchi}, \citenamefont
  {Watanabe}, \citenamefont {Smet},\ and\ \citenamefont
  {Zeldov}}]{Uri_WOS:000507726300002}%
  \BibitemOpen
  \bibfield  {author} {\bibinfo {author} {\bibfnamefont {A.}~\bibnamefont
  {Uri}}, \bibinfo {author} {\bibfnamefont {Y.}~\bibnamefont {Kim}}, \bibinfo
  {author} {\bibfnamefont {K.}~\bibnamefont {Bagani}}, \bibinfo {author}
  {\bibfnamefont {C.~K.}\ \bibnamefont {Lewandowski}}, \bibinfo {author}
  {\bibfnamefont {S.}~\bibnamefont {Grover}}, \bibinfo {author} {\bibfnamefont
  {N.}~\bibnamefont {Auerbach}}, \bibinfo {author} {\bibfnamefont {E.~O.}\
  \bibnamefont {Lachman}}, \bibinfo {author} {\bibfnamefont {Y.}~\bibnamefont
  {Myasoedov}}, \bibinfo {author} {\bibfnamefont {T.}~\bibnamefont
  {Taniguchi}}, \bibinfo {author} {\bibfnamefont {K.}~\bibnamefont {Watanabe}},
  \bibinfo {author} {\bibfnamefont {J.}~\bibnamefont {Smet}},\ and\ \bibinfo
  {author} {\bibfnamefont {E.}~\bibnamefont {Zeldov}},\ }\bibfield  {title}
  {\bibinfo {title} {{Nanoscale imaging of equilibrium quantum Hall edge
  currents and of the magnetic monopole response in graphene}},\ }\href
  {https://doi.org/10.1038/s41567-019-0713-3} {\bibfield  {journal} {\bibinfo
  {journal} {Nat. Phys.}\ }\textbf {\bibinfo {volume} {16}},\ \bibinfo {pages}
  {164} (\bibinfo {year} {2020})}\BibitemShut {NoStop}%
\bibitem [{\citenamefont {Johnsen}\ \emph {et~al.}(2023)\citenamefont
  {Johnsen}, \citenamefont {Schattauer}, \citenamefont {Samaddar},
  \citenamefont {Weston}, \citenamefont {Hamer}, \citenamefont {Watanabe},
  \citenamefont {Taniguchi}, \citenamefont {Gorbachev}, \citenamefont
  {Libisch},\ and\ \citenamefont {Morgenstern}}]{Johnsen_WOS:000961165700005}%
  \BibitemOpen
  \bibfield  {author} {\bibinfo {author} {\bibfnamefont {T.}~\bibnamefont
  {Johnsen}}, \bibinfo {author} {\bibfnamefont {C.}~\bibnamefont {Schattauer}},
  \bibinfo {author} {\bibfnamefont {S.}~\bibnamefont {Samaddar}}, \bibinfo
  {author} {\bibfnamefont {A.}~\bibnamefont {Weston}}, \bibinfo {author}
  {\bibfnamefont {M.~J.}\ \bibnamefont {Hamer}}, \bibinfo {author}
  {\bibfnamefont {K.}~\bibnamefont {Watanabe}}, \bibinfo {author}
  {\bibfnamefont {T.}~\bibnamefont {Taniguchi}}, \bibinfo {author}
  {\bibfnamefont {R.}~\bibnamefont {Gorbachev}}, \bibinfo {author}
  {\bibfnamefont {F.}~\bibnamefont {Libisch}},\ and\ \bibinfo {author}
  {\bibfnamefont {M.}~\bibnamefont {Morgenstern}},\ }\bibfield  {title}
  {\bibinfo {title} {{Mapping quantum Hall edge states in graphene by scanning
  tunneling microscopy}},\ }\href {https://doi.org/10.1103/PhysRevB.107.115426}
  {\bibfield  {journal} {\bibinfo  {journal} {Phys. Rev. B}\ }\textbf {\bibinfo
  {volume} {107}},\ \bibinfo {pages} {115426} (\bibinfo {year}
  {2023})}\BibitemShut {NoStop}%
\bibitem [{\citenamefont {Xu}\ \emph {et~al.}(2015)\citenamefont {Xu},
  \citenamefont {Belopolski}, \citenamefont {Sanchez}, \citenamefont {Zhang},
  \citenamefont {Chang}, \citenamefont {Guo}, \citenamefont {Bian},
  \citenamefont {Yuan}, \citenamefont {Lu}, \citenamefont {Chang},
  \citenamefont {Shibayev}, \citenamefont {Prokopovych}, \citenamefont
  {Alidoust}, \citenamefont {Zheng}, \citenamefont {Lee}, \citenamefont
  {Huang}, \citenamefont {Sankar}, \citenamefont {Chou}, \citenamefont {Hsu},
  \citenamefont {Jeng}, \citenamefont {Bansil}, \citenamefont {Neupert},
  \citenamefont {Strocov}, \citenamefont {Lin}, \citenamefont {Jia},\ and\
  \citenamefont {Hasan}}]{Xu_SciAdv_2015}%
  \BibitemOpen
  \bibfield  {author} {\bibinfo {author} {\bibfnamefont {S.-Y.}\ \bibnamefont
  {Xu}}, \bibinfo {author} {\bibfnamefont {I.}~\bibnamefont {Belopolski}},
  \bibinfo {author} {\bibfnamefont {D.~S.}\ \bibnamefont {Sanchez}}, \bibinfo
  {author} {\bibfnamefont {C.}~\bibnamefont {Zhang}}, \bibinfo {author}
  {\bibfnamefont {G.}~\bibnamefont {Chang}}, \bibinfo {author} {\bibfnamefont
  {C.}~\bibnamefont {Guo}}, \bibinfo {author} {\bibfnamefont {G.}~\bibnamefont
  {Bian}}, \bibinfo {author} {\bibfnamefont {Z.}~\bibnamefont {Yuan}}, \bibinfo
  {author} {\bibfnamefont {H.}~\bibnamefont {Lu}}, \bibinfo {author}
  {\bibfnamefont {T.-R.}\ \bibnamefont {Chang}}, \bibinfo {author}
  {\bibfnamefont {P.~P.}\ \bibnamefont {Shibayev}}, \bibinfo {author}
  {\bibfnamefont {M.~L.}\ \bibnamefont {Prokopovych}}, \bibinfo {author}
  {\bibfnamefont {N.}~\bibnamefont {Alidoust}}, \bibinfo {author}
  {\bibfnamefont {H.}~\bibnamefont {Zheng}}, \bibinfo {author} {\bibfnamefont
  {C.-C.}\ \bibnamefont {Lee}}, \bibinfo {author} {\bibfnamefont {S.-M.}\
  \bibnamefont {Huang}}, \bibinfo {author} {\bibfnamefont {R.}~\bibnamefont
  {Sankar}}, \bibinfo {author} {\bibfnamefont {F.}~\bibnamefont {Chou}},
  \bibinfo {author} {\bibfnamefont {C.-H.}\ \bibnamefont {Hsu}}, \bibinfo
  {author} {\bibfnamefont {H.-T.}\ \bibnamefont {Jeng}}, \bibinfo {author}
  {\bibfnamefont {A.}~\bibnamefont {Bansil}}, \bibinfo {author} {\bibfnamefont
  {T.}~\bibnamefont {Neupert}}, \bibinfo {author} {\bibfnamefont {V.~N.}\
  \bibnamefont {Strocov}}, \bibinfo {author} {\bibfnamefont {H.}~\bibnamefont
  {Lin}}, \bibinfo {author} {\bibfnamefont {S.}~\bibnamefont {Jia}},\ and\
  \bibinfo {author} {\bibfnamefont {M.~Z.}\ \bibnamefont {Hasan}},\ }\bibfield
  {title} {\bibinfo {title} {{Experimental discovery of a topological Weyl
  semimetal state in TaP}},\ }\href {https://doi.org/10.1126/sciadv.1501092}
  {\bibfield  {journal} {\bibinfo  {journal} {Sci. Adv.}\ }\textbf {\bibinfo
  {volume} {1}},\ \bibinfo {pages} {e1501092} (\bibinfo {year}
  {2015})}\BibitemShut {NoStop}%
\bibitem [{\citenamefont {Ferguson}\ \emph {et~al.}(2023)\citenamefont
  {Ferguson}, \citenamefont {Xiao}, \citenamefont {Richardella}, \citenamefont
  {Low}, \citenamefont {Samarth},\ and\ \citenamefont
  {Kowack}}]{Ferguson_2023}%
  \BibitemOpen
  \bibfield  {author} {\bibinfo {author} {\bibfnamefont {G.}~\bibnamefont
  {Ferguson}}, \bibinfo {author} {\bibfnamefont {R.}~\bibnamefont {Xiao}},
  \bibinfo {author} {\bibfnamefont {A.~R.}\ \bibnamefont {Richardella}},
  \bibinfo {author} {\bibfnamefont {D.}~\bibnamefont {Low}}, \bibinfo {author}
  {\bibfnamefont {N.}~\bibnamefont {Samarth}},\ and\ \bibinfo {author}
  {\bibfnamefont {K.~C.}\ \bibnamefont {Kowack}},\ }\bibfield  {title}
  {\bibinfo {title} {{Direct visualization of electronic transport in a quantum
  anomalous Hall insulator}},\ }\href
  {https://doi.org/10.1038/s41563-023-01622-0} {\bibfield  {journal} {\bibinfo
  {journal} {Nat. Mater.}\ }\textbf {\bibinfo {volume} {22}},\ \bibinfo {pages}
  {1100} (\bibinfo {year} {2023})}\BibitemShut {NoStop}%
\bibitem [{\citenamefont {\ifmmode~\mbox{\c{S}}\else \c{S}\fi{}ahin}\ and\
  \citenamefont {Flatt\'e}(2015)}]{Sahin2015}%
  \BibitemOpen
  \bibfield  {author} {\bibinfo {author} {\bibfnamefont {C.}~\bibnamefont
  {\ifmmode~\mbox{\c{S}}\else \c{S}\fi{}ahin}}\ and\ \bibinfo {author}
  {\bibfnamefont {M.~E.}\ \bibnamefont {Flatt\'e}},\ }\bibfield  {title}
  {\bibinfo {title} {Tunable giant spin {Hall} conductivities in a strong
  spin-orbit semimetal:
  {${\mathrm{Bi}}_{1\ensuremath{-}x}{\mathrm{Sb}}_{x}$}},\ }\href
  {https://doi.org/10.1103/PhysRevLett.114.107201} {\bibfield  {journal}
  {\bibinfo  {journal} {Phys. Rev. Lett.}\ }\textbf {\bibinfo {volume} {114}},\
  \bibinfo {pages} {107201} (\bibinfo {year} {2015})}\BibitemShut {NoStop}%
\bibitem [{\citenamefont {Khang}\ \emph {et~al.}(2018)\citenamefont {Khang},
  \citenamefont {Ueda},\ and\ \citenamefont {Hai}}]{Khang2018}%
  \BibitemOpen
  \bibfield  {author} {\bibinfo {author} {\bibfnamefont {N.~H.~D.}\
  \bibnamefont {Khang}}, \bibinfo {author} {\bibfnamefont {Y.}~\bibnamefont
  {Ueda}},\ and\ \bibinfo {author} {\bibfnamefont {P.~N.}\ \bibnamefont
  {Hai}},\ }\bibfield  {title} {\bibinfo {title} {A conductive topological
  insulator with large spin {H}all effect for ultralow power spin--orbit torque
  switching},\ }\href {https://doi.org/10.1038/s41563-018-0137-y} {\bibfield
  {journal} {\bibinfo  {journal} {Nat. Mater.}\ }\textbf {\bibinfo {volume}
  {17}},\ \bibinfo {pages} {808} (\bibinfo {year} {2018})}\BibitemShut
  {NoStop}%
\bibitem [{\citenamefont {Roschewsky}\ \emph {et~al.}(2019)\citenamefont
  {Roschewsky}, \citenamefont {Walker}, \citenamefont {Gowtham}, \citenamefont
  {Muschinske}, \citenamefont {Hellman}, \citenamefont {Bank},\ and\
  \citenamefont {Salahuddin}}]{Roschewsky2019}%
  \BibitemOpen
  \bibfield  {author} {\bibinfo {author} {\bibfnamefont {N.}~\bibnamefont
  {Roschewsky}}, \bibinfo {author} {\bibfnamefont {E.~S.}\ \bibnamefont
  {Walker}}, \bibinfo {author} {\bibfnamefont {P.}~\bibnamefont {Gowtham}},
  \bibinfo {author} {\bibfnamefont {S.}~\bibnamefont {Muschinske}}, \bibinfo
  {author} {\bibfnamefont {F.}~\bibnamefont {Hellman}}, \bibinfo {author}
  {\bibfnamefont {S.~R.}\ \bibnamefont {Bank}},\ and\ \bibinfo {author}
  {\bibfnamefont {S.}~\bibnamefont {Salahuddin}},\ }\bibfield  {title}
  {\bibinfo {title} {Spin-orbit torque and {N}ernst effect in {Bi-Sb/Co}
  heterostructures},\ }\href
  {https://journals.aps.org/prb/abstract/10.1103/PhysRevB.99.195103} {\bibfield
   {journal} {\bibinfo  {journal} {Phys. Rev. B}\ }\textbf {\bibinfo {volume}
  {99}},\ \bibinfo {pages} {195103} (\bibinfo {year} {2019})}\BibitemShut
  {NoStop}%
\bibitem [{\citenamefont {Khang}\ \emph {et~al.}(2020)\citenamefont {Khang},
  \citenamefont {Nakano}, \citenamefont {Shirokura}, \citenamefont {Miyamoto},\
  and\ \citenamefont {Hai}}]{Khang2020}%
  \BibitemOpen
  \bibfield  {author} {\bibinfo {author} {\bibfnamefont {N.~H.~D.}\
  \bibnamefont {Khang}}, \bibinfo {author} {\bibfnamefont {S.}~\bibnamefont
  {Nakano}}, \bibinfo {author} {\bibfnamefont {T.}~\bibnamefont {Shirokura}},
  \bibinfo {author} {\bibfnamefont {Y.}~\bibnamefont {Miyamoto}},\ and\
  \bibinfo {author} {\bibfnamefont {P.~N.}\ \bibnamefont {Hai}},\ }\bibfield
  {title} {\bibinfo {title} {Ultralow power spin--orbit torque magnetization
  switching induced by a non-epitaxial topological insulator on {Si}
  substrates},\ }\href {https://doi.org/10.1038/s41598-020-69027-6} {\bibfield
  {journal} {\bibinfo  {journal} {Sci. Rep.}\ }\textbf {\bibinfo {volume}
  {10}},\ \bibinfo {pages} {12185} (\bibinfo {year} {2020})}\BibitemShut
  {NoStop}%
\bibitem [{\citenamefont {Chi}\ \emph {et~al.}(2020)\citenamefont {Chi},
  \citenamefont {Lau}, \citenamefont {Xu}, \citenamefont {Ohkubo},
  \citenamefont {Hono},\ and\ \citenamefont {Hayashi}}]{Chi2020}%
  \BibitemOpen
  \bibfield  {author} {\bibinfo {author} {\bibfnamefont {Z.}~\bibnamefont
  {Chi}}, \bibinfo {author} {\bibfnamefont {Y.~C.}\ \bibnamefont {Lau}},
  \bibinfo {author} {\bibfnamefont {X.}~\bibnamefont {Xu}}, \bibinfo {author}
  {\bibfnamefont {T.}~\bibnamefont {Ohkubo}}, \bibinfo {author} {\bibfnamefont
  {K.}~\bibnamefont {Hono}},\ and\ \bibinfo {author} {\bibfnamefont
  {M.}~\bibnamefont {Hayashi}},\ }\bibfield  {title} {\bibinfo {title} {{The
  spin {H}all effect of {Bi-Sb} alloys driven by thermally excited {D}irac-like
  electrons}},\ }\href {https://doi.org/10.1126/sciadv.aay2324} {\bibfield
  {journal} {\bibinfo  {journal} {Sci. Adv.}\ }\textbf {\bibinfo {volume}
  {6}},\ \bibinfo {pages} {aay2324} (\bibinfo {year} {2020})}\BibitemShut
  {NoStop}%
\bibitem [{\citenamefont {Liu}\ \emph {et~al.}(2011)\citenamefont {Liu},
  \citenamefont {Moriyama}, \citenamefont {Ralph},\ and\ \citenamefont
  {Buhrman}}]{PhysRevLett.106.036601LiuSTFMR}%
  \BibitemOpen
  \bibfield  {author} {\bibinfo {author} {\bibfnamefont {L.}~\bibnamefont
  {Liu}}, \bibinfo {author} {\bibfnamefont {T.}~\bibnamefont {Moriyama}},
  \bibinfo {author} {\bibfnamefont {D.~C.}\ \bibnamefont {Ralph}},\ and\
  \bibinfo {author} {\bibfnamefont {R.~A.}\ \bibnamefont {Buhrman}},\
  }\bibfield  {title} {\bibinfo {title} {Spin-torque ferromagnetic resonance
  induced by the spin {H}all effect},\ }\href
  {https://doi.org/10.1103/PhysRevLett.106.036601} {\bibfield  {journal}
  {\bibinfo  {journal} {Phys. Rev. Lett.}\ }\textbf {\bibinfo {volume} {106}},\
  \bibinfo {pages} {036601} (\bibinfo {year} {2011})}\BibitemShut {NoStop}%
\bibitem [{\citenamefont {Wang}\ \emph {et~al.}(2019)\citenamefont {Wang},
  \citenamefont {Kally}, \citenamefont {\ifmmode~\mbox{\c{S}}\else
  \c{S}\fi{}ahin}, \citenamefont {Liu}, \citenamefont {Yanez}, \citenamefont
  {Kamp}, \citenamefont {Richardella}, \citenamefont {Wu}, \citenamefont
  {Flatt\'e},\ and\ \citenamefont {Samarth}}]{HWang_PhysRevResearch.1.012014}%
  \BibitemOpen
  \bibfield  {author} {\bibinfo {author} {\bibfnamefont {H.}~\bibnamefont
  {Wang}}, \bibinfo {author} {\bibfnamefont {J.}~\bibnamefont {Kally}},
  \bibinfo {author} {\bibfnamefont {C.}~\bibnamefont
  {\ifmmode~\mbox{\c{S}}\else \c{S}\fi{}ahin}}, \bibinfo {author}
  {\bibfnamefont {T.}~\bibnamefont {Liu}}, \bibinfo {author} {\bibfnamefont
  {W.}~\bibnamefont {Yanez}}, \bibinfo {author} {\bibfnamefont {E.~J.}\
  \bibnamefont {Kamp}}, \bibinfo {author} {\bibfnamefont {A.}~\bibnamefont
  {Richardella}}, \bibinfo {author} {\bibfnamefont {M.}~\bibnamefont {Wu}},
  \bibinfo {author} {\bibfnamefont {M.~E.}\ \bibnamefont {Flatt\'e}},\ and\
  \bibinfo {author} {\bibfnamefont {N.}~\bibnamefont {Samarth}},\ }\bibfield
  {title} {\bibinfo {title} {Fermi level dependent spin pumping from a magnetic
  insulator into a topological insulator},\ }\href
  {https://doi.org/10.1103/PhysRevResearch.1.012014} {\bibfield  {journal}
  {\bibinfo  {journal} {Phys. Rev. Research}\ }\textbf {\bibinfo {volume}
  {1}},\ \bibinfo {pages} {012014(R)} (\bibinfo {year} {2019})}\BibitemShut
  {NoStop}%
\bibitem [{\citenamefont {Hellman}\ \emph {et~al.}(2017)\citenamefont
  {Hellman}, \citenamefont {Hoffmann}, \citenamefont {Tserkovnyak},
  \citenamefont {Beach}, \citenamefont {Fullerton}, \citenamefont {Leighton},
  \citenamefont {MacDonald}, \citenamefont {Ralph}, \citenamefont {Arena},
  \citenamefont {D\"urr}, \citenamefont {Fischer}, \citenamefont {Grollier},
  \citenamefont {Heremans}, \citenamefont {Jungwirth}, \citenamefont {Kimel},
  \citenamefont {Koopmans}, \citenamefont {Krivorotov}, \citenamefont {May},
  \citenamefont {Petford-Long}, \citenamefont {Rondinelli}, \citenamefont
  {Samarth}, \citenamefont {Schuller}, \citenamefont {Slavin}, \citenamefont
  {Stiles}, \citenamefont {Tchernyshyov}, \citenamefont {Thiaville},\ and\
  \citenamefont {Zink}}]{Hellman_RevModPhys.89.025006}%
  \BibitemOpen
  \bibfield  {author} {\bibinfo {author} {\bibfnamefont {F.}~\bibnamefont
  {Hellman}}, \bibinfo {author} {\bibfnamefont {A.}~\bibnamefont {Hoffmann}},
  \bibinfo {author} {\bibfnamefont {Y.}~\bibnamefont {Tserkovnyak}}, \bibinfo
  {author} {\bibfnamefont {G.~S.~D.}\ \bibnamefont {Beach}}, \bibinfo {author}
  {\bibfnamefont {E.~E.}\ \bibnamefont {Fullerton}}, \bibinfo {author}
  {\bibfnamefont {C.}~\bibnamefont {Leighton}}, \bibinfo {author}
  {\bibfnamefont {A.~H.}\ \bibnamefont {MacDonald}}, \bibinfo {author}
  {\bibfnamefont {D.~C.}\ \bibnamefont {Ralph}}, \bibinfo {author}
  {\bibfnamefont {D.~A.}\ \bibnamefont {Arena}}, \bibinfo {author}
  {\bibfnamefont {H.~A.}\ \bibnamefont {D\"urr}}, \bibinfo {author}
  {\bibfnamefont {P.}~\bibnamefont {Fischer}}, \bibinfo {author} {\bibfnamefont
  {J.}~\bibnamefont {Grollier}}, \bibinfo {author} {\bibfnamefont {J.~P.}\
  \bibnamefont {Heremans}}, \bibinfo {author} {\bibfnamefont {T.}~\bibnamefont
  {Jungwirth}}, \bibinfo {author} {\bibfnamefont {A.~V.}\ \bibnamefont
  {Kimel}}, \bibinfo {author} {\bibfnamefont {B.}~\bibnamefont {Koopmans}},
  \bibinfo {author} {\bibfnamefont {I.~N.}\ \bibnamefont {Krivorotov}},
  \bibinfo {author} {\bibfnamefont {S.~J.}\ \bibnamefont {May}}, \bibinfo
  {author} {\bibfnamefont {A.~K.}\ \bibnamefont {Petford-Long}}, \bibinfo
  {author} {\bibfnamefont {J.~M.}\ \bibnamefont {Rondinelli}}, \bibinfo
  {author} {\bibfnamefont {N.}~\bibnamefont {Samarth}}, \bibinfo {author}
  {\bibfnamefont {I.~K.}\ \bibnamefont {Schuller}}, \bibinfo {author}
  {\bibfnamefont {A.~N.}\ \bibnamefont {Slavin}}, \bibinfo {author}
  {\bibfnamefont {M.~D.}\ \bibnamefont {Stiles}}, \bibinfo {author}
  {\bibfnamefont {O.}~\bibnamefont {Tchernyshyov}}, \bibinfo {author}
  {\bibfnamefont {A.}~\bibnamefont {Thiaville}},\ and\ \bibinfo {author}
  {\bibfnamefont {B.~L.}\ \bibnamefont {Zink}},\ }\bibfield  {title} {\bibinfo
  {title} {Interface-induced phenomena in magnetism},\ }\href
  {https://doi.org/10.1103/RevModPhys.89.025006} {\bibfield  {journal}
  {\bibinfo  {journal} {Rev. Mod. Phys.}\ }\textbf {\bibinfo {volume} {89}},\
  \bibinfo {pages} {025006} (\bibinfo {year} {2017})}\BibitemShut {NoStop}%
\bibitem [{\citenamefont {Manchon}\ \emph {et~al.}(2019)\citenamefont
  {Manchon}, \citenamefont {\ifmmode~\check{Z}\else \v{Z}\fi{}elezn\'y},
  \citenamefont {Miron}, \citenamefont {Jungwirth}, \citenamefont {Sinova},
  \citenamefont {Thiaville}, \citenamefont {Garello},\ and\ \citenamefont
  {Gambardella}}]{Miron_RevModPhys.91.035004}%
  \BibitemOpen
  \bibfield  {author} {\bibinfo {author} {\bibfnamefont {A.}~\bibnamefont
  {Manchon}}, \bibinfo {author} {\bibfnamefont {J.}~\bibnamefont
  {\ifmmode~\check{Z}\else \v{Z}\fi{}elezn\'y}}, \bibinfo {author}
  {\bibfnamefont {I.~M.}\ \bibnamefont {Miron}}, \bibinfo {author}
  {\bibfnamefont {T.}~\bibnamefont {Jungwirth}}, \bibinfo {author}
  {\bibfnamefont {J.}~\bibnamefont {Sinova}}, \bibinfo {author} {\bibfnamefont
  {A.}~\bibnamefont {Thiaville}}, \bibinfo {author} {\bibfnamefont
  {K.}~\bibnamefont {Garello}},\ and\ \bibinfo {author} {\bibfnamefont
  {P.}~\bibnamefont {Gambardella}},\ }\bibfield  {title} {\bibinfo {title}
  {Current-induced spin-orbit torques in ferromagnetic and antiferromagnetic
  systems},\ }\href {https://doi.org/10.1103/RevModPhys.91.035004} {\bibfield
  {journal} {\bibinfo  {journal} {Rev. Mod. Phys.}\ }\textbf {\bibinfo {volume}
  {91}},\ \bibinfo {pages} {035004} (\bibinfo {year} {2019})}\BibitemShut
  {NoStop}%
\bibitem [{\citenamefont {Miron}\ \emph {et~al.}(2010)\citenamefont {Miron},
  \citenamefont {Gaudin}, \citenamefont {Auffret}, \citenamefont {Rodmacq},
  \citenamefont {Schuhl}, \citenamefont {Pizzini}, \citenamefont {Vogel},\ and\
  \citenamefont {P.}}]{Miron_NatMat}%
  \BibitemOpen
  \bibfield  {author} {\bibinfo {author} {\bibfnamefont {I.~M.}\ \bibnamefont
  {Miron}}, \bibinfo {author} {\bibfnamefont {G.}~\bibnamefont {Gaudin}},
  \bibinfo {author} {\bibfnamefont {S.}~\bibnamefont {Auffret}}, \bibinfo
  {author} {\bibfnamefont {B.}~\bibnamefont {Rodmacq}}, \bibinfo {author}
  {\bibfnamefont {A.}~\bibnamefont {Schuhl}}, \bibinfo {author} {\bibfnamefont
  {S.}~\bibnamefont {Pizzini}}, \bibinfo {author} {\bibfnamefont
  {J.}~\bibnamefont {Vogel}},\ and\ \bibinfo {author} {\bibfnamefont
  {G.}~\bibnamefont {P.}},\ }\bibfield  {title} {\bibinfo {title}
  {Current-driven spin torque induced by the {Rashba} effect in a ferromagnetic
  metal layer},\ }\href {https://www.nature.com/articles/nmat2613} {\bibfield
  {journal} {\bibinfo  {journal} {Nat. Mater}\ }\textbf {\bibinfo {volume}
  {9}},\ \bibinfo {pages} {230} (\bibinfo {year} {2010})}\BibitemShut {NoStop}%
\bibitem [{\citenamefont {Liu}\ \emph {et~al.}(2012)\citenamefont {Liu},
  \citenamefont {Pai}, \citenamefont {Li}, \citenamefont {Tseng}, \citenamefont
  {Ralph},\ and\ \citenamefont {Buhrman}}]{Liu2012}%
  \BibitemOpen
  \bibfield  {author} {\bibinfo {author} {\bibfnamefont {L.}~\bibnamefont
  {Liu}}, \bibinfo {author} {\bibfnamefont {C.~F.}\ \bibnamefont {Pai}},
  \bibinfo {author} {\bibfnamefont {Y.}~\bibnamefont {Li}}, \bibinfo {author}
  {\bibfnamefont {H.~W.}\ \bibnamefont {Tseng}}, \bibinfo {author}
  {\bibfnamefont {D.~C.}\ \bibnamefont {Ralph}},\ and\ \bibinfo {author}
  {\bibfnamefont {R.~A.}\ \bibnamefont {Buhrman}},\ }\bibfield  {title}
  {\bibinfo {title} {Spin-torque switching with the giant spin {H}all effect of
  tantalum},\ }\href {https://doi.org/10.1126/science.1218197} {\bibfield
  {journal} {\bibinfo  {journal} {Science}\ }\textbf {\bibinfo {volume}
  {336}},\ \bibinfo {pages} {555} (\bibinfo {year} {2012})}\BibitemShut
  {NoStop}%
\bibitem [{\citenamefont {Mellnik}\ \emph {et~al.}(2014)\citenamefont
  {Mellnik}, \citenamefont {Lee}, \citenamefont {Richardella}, \citenamefont
  {Grab}, \citenamefont {Mintun}, \citenamefont {Fischer}, \citenamefont
  {Vaezi}, \citenamefont {Manchon}, \citenamefont {Kim}, \citenamefont
  {Samarth},\ and\ \citenamefont {Ralph}}]{Mellnik2014}%
  \BibitemOpen
  \bibfield  {author} {\bibinfo {author} {\bibfnamefont {A.~R.}\ \bibnamefont
  {Mellnik}}, \bibinfo {author} {\bibfnamefont {J.~S.}\ \bibnamefont {Lee}},
  \bibinfo {author} {\bibfnamefont {A.}~\bibnamefont {Richardella}}, \bibinfo
  {author} {\bibfnamefont {J.~L.}\ \bibnamefont {Grab}}, \bibinfo {author}
  {\bibfnamefont {P.~J.}\ \bibnamefont {Mintun}}, \bibinfo {author}
  {\bibfnamefont {M.~H.}\ \bibnamefont {Fischer}}, \bibinfo {author}
  {\bibfnamefont {A.}~\bibnamefont {Vaezi}}, \bibinfo {author} {\bibfnamefont
  {A.}~\bibnamefont {Manchon}}, \bibinfo {author} {\bibfnamefont {E.-A.}\
  \bibnamefont {Kim}}, \bibinfo {author} {\bibfnamefont {N.}~\bibnamefont
  {Samarth}},\ and\ \bibinfo {author} {\bibfnamefont {D.~C.}\ \bibnamefont
  {Ralph}},\ }\bibfield  {title} {\bibinfo {title} {Spin-transfer torque
  generated by a topological insulator},\ }\href
  {https://doi.org/10.1038/nature13534} {\bibfield  {journal} {\bibinfo
  {journal} {Nature (London)}\ }\textbf {\bibinfo {volume} {511}},\ \bibinfo
  {pages} {449} (\bibinfo {year} {2014})}\BibitemShut {NoStop}%
\bibitem [{\citenamefont {Kondou}\ \emph {et~al.}(2016)\citenamefont {Kondou},
  \citenamefont {Yoshimi}, \citenamefont {Tsukazaki}, \citenamefont {Fukuma},
  \citenamefont {Matsuno}, \citenamefont {Takahashi}, \citenamefont {Kawasaki},
  \citenamefont {Tokura},\ and\ \citenamefont {Otani}}]{Kondou2016}%
  \BibitemOpen
  \bibfield  {author} {\bibinfo {author} {\bibfnamefont {K.}~\bibnamefont
  {Kondou}}, \bibinfo {author} {\bibfnamefont {R.}~\bibnamefont {Yoshimi}},
  \bibinfo {author} {\bibfnamefont {A.}~\bibnamefont {Tsukazaki}}, \bibinfo
  {author} {\bibfnamefont {Y.}~\bibnamefont {Fukuma}}, \bibinfo {author}
  {\bibfnamefont {J.}~\bibnamefont {Matsuno}}, \bibinfo {author} {\bibfnamefont
  {K.~S.}\ \bibnamefont {Takahashi}}, \bibinfo {author} {\bibfnamefont
  {M.}~\bibnamefont {Kawasaki}}, \bibinfo {author} {\bibfnamefont
  {Y.}~\bibnamefont {Tokura}},\ and\ \bibinfo {author} {\bibfnamefont
  {Y.}~\bibnamefont {Otani}},\ }\bibfield  {title} {\bibinfo {title}
  {Fermi-level-dependent charge-to-spin current conversion by {D}irac surface
  states of topological insulators},\ }\href
  {https://doi.org/10.1038/nphys3833} {\bibfield  {journal} {\bibinfo
  {journal} {Nat. Phys.}\ }\textbf {\bibinfo {volume} {12}},\ \bibinfo {pages}
  {1027} (\bibinfo {year} {2016})}\BibitemShut {NoStop}%
\bibitem [{\citenamefont {DC}\ \emph {et~al.}(2018)\citenamefont {DC},
  \citenamefont {Grassi}, \citenamefont {Chen}, \citenamefont {Jamali},
  \citenamefont {Reifsnyder~Hickey}, \citenamefont {Zhang}, \citenamefont
  {Zhao}, \citenamefont {Li}, \citenamefont {Quarterman}, \citenamefont {Lv},
  \citenamefont {Li}, \citenamefont {Manchon}, \citenamefont {Mkhoyan},
  \citenamefont {Low},\ and\ \citenamefont {Wang}}]{DC2018}%
  \BibitemOpen
  \bibfield  {author} {\bibinfo {author} {\bibfnamefont {M.}~\bibnamefont
  {DC}}, \bibinfo {author} {\bibfnamefont {R.}~\bibnamefont {Grassi}}, \bibinfo
  {author} {\bibfnamefont {J.-Y.}\ \bibnamefont {Chen}}, \bibinfo {author}
  {\bibfnamefont {M.}~\bibnamefont {Jamali}}, \bibinfo {author} {\bibfnamefont
  {D.}~\bibnamefont {Reifsnyder~Hickey}}, \bibinfo {author} {\bibfnamefont
  {D.}~\bibnamefont {Zhang}}, \bibinfo {author} {\bibfnamefont
  {Z.}~\bibnamefont {Zhao}}, \bibinfo {author} {\bibfnamefont {H.}~\bibnamefont
  {Li}}, \bibinfo {author} {\bibfnamefont {P.}~\bibnamefont {Quarterman}},
  \bibinfo {author} {\bibfnamefont {Y.}~\bibnamefont {Lv}}, \bibinfo {author}
  {\bibfnamefont {M.}~\bibnamefont {Li}}, \bibinfo {author} {\bibfnamefont
  {A.}~\bibnamefont {Manchon}}, \bibinfo {author} {\bibfnamefont {K.~A.}\
  \bibnamefont {Mkhoyan}}, \bibinfo {author} {\bibfnamefont {T.}~\bibnamefont
  {Low}},\ and\ \bibinfo {author} {\bibfnamefont {J.-P.}\ \bibnamefont
  {Wang}},\ }\bibfield  {title} {\bibinfo {title} {Room-temperature high
  spin--orbit torque due to quantum confinement in sputtered
  $\mathrm{Bi_xSe_{(1-x)}}$ films},\ }\href
  {https://doi.org/10.1038/s41563-018-0136-z} {\bibfield  {journal} {\bibinfo
  {journal} {Nat. Mater.}\ }\textbf {\bibinfo {volume} {17}},\ \bibinfo {pages}
  {800} (\bibinfo {year} {2018})}\BibitemShut {NoStop}%
\bibitem [{\citenamefont {Wu}\ \emph {et~al.}(2019)\citenamefont {Wu},
  \citenamefont {Zhang}, \citenamefont {Deng}, \citenamefont {Lan},
  \citenamefont {Pan}, \citenamefont {Razavi}, \citenamefont {Che},
  \citenamefont {Huang}, \citenamefont {Dai}, \citenamefont {Wong},
  \citenamefont {Han},\ and\ \citenamefont {Wang}}]{PhysRevLett.123.207205}%
  \BibitemOpen
  \bibfield  {author} {\bibinfo {author} {\bibfnamefont {H.}~\bibnamefont
  {Wu}}, \bibinfo {author} {\bibfnamefont {P.}~\bibnamefont {Zhang}}, \bibinfo
  {author} {\bibfnamefont {P.}~\bibnamefont {Deng}}, \bibinfo {author}
  {\bibfnamefont {Q.}~\bibnamefont {Lan}}, \bibinfo {author} {\bibfnamefont
  {Q.}~\bibnamefont {Pan}}, \bibinfo {author} {\bibfnamefont {S.~A.}\
  \bibnamefont {Razavi}}, \bibinfo {author} {\bibfnamefont {X.}~\bibnamefont
  {Che}}, \bibinfo {author} {\bibfnamefont {L.}~\bibnamefont {Huang}}, \bibinfo
  {author} {\bibfnamefont {B.}~\bibnamefont {Dai}}, \bibinfo {author}
  {\bibfnamefont {K.}~\bibnamefont {Wong}}, \bibinfo {author} {\bibfnamefont
  {X.}~\bibnamefont {Han}},\ and\ \bibinfo {author} {\bibfnamefont {K.~L.}\
  \bibnamefont {Wang}},\ }\bibfield  {title} {\bibinfo {title}
  {Room-temperature spin-orbit torque from topological surface states},\ }\href
  {https://doi.org/10.1103/PhysRevLett.123.207205} {\bibfield  {journal}
  {\bibinfo  {journal} {Phys. Rev. Lett.}\ }\textbf {\bibinfo {volume} {123}},\
  \bibinfo {pages} {207205} (\bibinfo {year} {2019})}\BibitemShut {NoStop}%
\bibitem [{\citenamefont {Yanez}\ \emph {et~al.}(2021)\citenamefont {Yanez},
  \citenamefont {Ou}, \citenamefont {Xiao}, \citenamefont {Koo}, \citenamefont
  {Held}, \citenamefont {Ghosh}, \citenamefont {Rable}, \citenamefont
  {Pillsbury}, \citenamefont {Delgado}, \citenamefont {Yang}, \citenamefont
  {Chamorro}, \citenamefont {Grutter}, \citenamefont {Quarterman},
  \citenamefont {Richardella}, \citenamefont {Sengupta}, \citenamefont
  {McQueen}, \citenamefont {Borchers}, \citenamefont {Mkhoyan}, \citenamefont
  {Yan},\ and\ \citenamefont {Samarth}}]{PhysRevApplied.16.054031yanez}%
  \BibitemOpen
  \bibfield  {author} {\bibinfo {author} {\bibfnamefont {W.}~\bibnamefont
  {Yanez}}, \bibinfo {author} {\bibfnamefont {Y.}~\bibnamefont {Ou}}, \bibinfo
  {author} {\bibfnamefont {R.}~\bibnamefont {Xiao}}, \bibinfo {author}
  {\bibfnamefont {J.}~\bibnamefont {Koo}}, \bibinfo {author} {\bibfnamefont
  {J.~T.}\ \bibnamefont {Held}}, \bibinfo {author} {\bibfnamefont
  {S.}~\bibnamefont {Ghosh}}, \bibinfo {author} {\bibfnamefont
  {J.}~\bibnamefont {Rable}}, \bibinfo {author} {\bibfnamefont
  {T.}~\bibnamefont {Pillsbury}}, \bibinfo {author} {\bibfnamefont {E.~G.}\
  \bibnamefont {Delgado}}, \bibinfo {author} {\bibfnamefont {K.}~\bibnamefont
  {Yang}}, \bibinfo {author} {\bibfnamefont {J.}~\bibnamefont {Chamorro}},
  \bibinfo {author} {\bibfnamefont {A.~J.}\ \bibnamefont {Grutter}}, \bibinfo
  {author} {\bibfnamefont {P.}~\bibnamefont {Quarterman}}, \bibinfo {author}
  {\bibfnamefont {A.}~\bibnamefont {Richardella}}, \bibinfo {author}
  {\bibfnamefont {A.}~\bibnamefont {Sengupta}}, \bibinfo {author}
  {\bibfnamefont {T.}~\bibnamefont {McQueen}}, \bibinfo {author} {\bibfnamefont
  {J.~A.}\ \bibnamefont {Borchers}}, \bibinfo {author} {\bibfnamefont {K.~A.}\
  \bibnamefont {Mkhoyan}}, \bibinfo {author} {\bibfnamefont {B.}~\bibnamefont
  {Yan}},\ and\ \bibinfo {author} {\bibfnamefont {N.}~\bibnamefont {Samarth}},\
  }\bibfield  {title} {\bibinfo {title} {Spin and charge interconversion in
  {D}irac-semimetal thin films},\ }\href
  {https://link.aps.org/doi/10.1103/PhysRevApplied.16.054031} {\bibfield
  {journal} {\bibinfo  {journal} {Phys. Rev. Appl.}\ }\textbf {\bibinfo
  {volume} {16}},\ \bibinfo {pages} {054031} (\bibinfo {year}
  {2021})}\BibitemShut {NoStop}%
\bibitem [{\citenamefont {Yanez}\ \emph {et~al.}(2022)\citenamefont {Yanez},
  \citenamefont {Ou}, \citenamefont {Xiao}, \citenamefont {Ghosh},
  \citenamefont {Dwivedi}, \citenamefont {Steinebronn}, \citenamefont
  {Richardella}, \citenamefont {Mkhoyan},\ and\ \citenamefont
  {Samarth}}]{Yanez_PhysRevApplied.18.054004}%
  \BibitemOpen
  \bibfield  {author} {\bibinfo {author} {\bibfnamefont {W.}~\bibnamefont
  {Yanez}}, \bibinfo {author} {\bibfnamefont {Y.}~\bibnamefont {Ou}}, \bibinfo
  {author} {\bibfnamefont {R.}~\bibnamefont {Xiao}}, \bibinfo {author}
  {\bibfnamefont {S.}~\bibnamefont {Ghosh}}, \bibinfo {author} {\bibfnamefont
  {J.}~\bibnamefont {Dwivedi}}, \bibinfo {author} {\bibfnamefont
  {E.}~\bibnamefont {Steinebronn}}, \bibinfo {author} {\bibfnamefont
  {A.}~\bibnamefont {Richardella}}, \bibinfo {author} {\bibfnamefont {K.~A.}\
  \bibnamefont {Mkhoyan}},\ and\ \bibinfo {author} {\bibfnamefont
  {N.}~\bibnamefont {Samarth}},\ }\bibfield  {title} {\bibinfo {title} {Giant
  dampinglike-torque efficiency in naturally oxidized polycrystalline
  $\mathrm{Ta}\mathrm{As}$ thin films},\ }\href
  {https://doi.org/10.1103/PhysRevApplied.18.054004} {\bibfield  {journal}
  {\bibinfo  {journal} {Phys. Rev. Appl.}\ }\textbf {\bibinfo {volume} {18}},\
  \bibinfo {pages} {054004} (\bibinfo {year} {2022})}\BibitemShut {NoStop}%
\bibitem [{\citenamefont {Fan}\ \emph {et~al.}(2014)\citenamefont {Fan},
  \citenamefont {Upadhyaya}, \citenamefont {Kou}, \citenamefont {Lang},
  \citenamefont {Takei}, \citenamefont {Wang}, \citenamefont {Tang},
  \citenamefont {He}, \citenamefont {Chang}, \citenamefont {Montazeri},
  \citenamefont {Yu}, \citenamefont {Jiang}, \citenamefont {Nie}, \citenamefont
  {Schwartz}, \citenamefont {Tserkovnyak},\ and\ \citenamefont
  {Wang}}]{Fan_NMat_2014}%
  \BibitemOpen
  \bibfield  {author} {\bibinfo {author} {\bibfnamefont {Y.}~\bibnamefont
  {Fan}}, \bibinfo {author} {\bibfnamefont {P.}~\bibnamefont {Upadhyaya}},
  \bibinfo {author} {\bibfnamefont {X.}~\bibnamefont {Kou}}, \bibinfo {author}
  {\bibfnamefont {M.}~\bibnamefont {Lang}}, \bibinfo {author} {\bibfnamefont
  {S.}~\bibnamefont {Takei}}, \bibinfo {author} {\bibfnamefont
  {Z.}~\bibnamefont {Wang}}, \bibinfo {author} {\bibfnamefont {J.}~\bibnamefont
  {Tang}}, \bibinfo {author} {\bibfnamefont {L.}~\bibnamefont {He}}, \bibinfo
  {author} {\bibfnamefont {L.-T.}\ \bibnamefont {Chang}}, \bibinfo {author}
  {\bibfnamefont {M.}~\bibnamefont {Montazeri}}, \bibinfo {author}
  {\bibfnamefont {G.}~\bibnamefont {Yu}}, \bibinfo {author} {\bibfnamefont
  {W.}~\bibnamefont {Jiang}}, \bibinfo {author} {\bibfnamefont
  {T.}~\bibnamefont {Nie}}, \bibinfo {author} {\bibfnamefont {R.~N.}\
  \bibnamefont {Schwartz}}, \bibinfo {author} {\bibfnamefont {Y.}~\bibnamefont
  {Tserkovnyak}},\ and\ \bibinfo {author} {\bibfnamefont {K.~L.}\ \bibnamefont
  {Wang}},\ }\bibfield  {title} {\bibinfo {title} {Magnetization switching
  through giant spin--orbit torque in a magnetically doped topological
  insulator heterostructure},\ }\href {https://doi.org/10.1038/nmat3973}
  {\bibfield  {journal} {\bibinfo  {journal} {Nat. Mater.}\ }\textbf {\bibinfo
  {volume} {13}},\ \bibinfo {pages} {699} (\bibinfo {year} {2014})}\BibitemShut
  {NoStop}%
\bibitem [{\citenamefont {Han}\ \emph {et~al.}(2017)\citenamefont {Han},
  \citenamefont {Richardella}, \citenamefont {Siddiqui}, \citenamefont
  {Finley}, \citenamefont {Samarth},\ and\ \citenamefont
  {Liu}}]{Han_PhysRevLett.119.077702}%
  \BibitemOpen
  \bibfield  {author} {\bibinfo {author} {\bibfnamefont {J.}~\bibnamefont
  {Han}}, \bibinfo {author} {\bibfnamefont {A.}~\bibnamefont {Richardella}},
  \bibinfo {author} {\bibfnamefont {S.~A.}\ \bibnamefont {Siddiqui}}, \bibinfo
  {author} {\bibfnamefont {J.}~\bibnamefont {Finley}}, \bibinfo {author}
  {\bibfnamefont {N.}~\bibnamefont {Samarth}},\ and\ \bibinfo {author}
  {\bibfnamefont {L.}~\bibnamefont {Liu}},\ }\bibfield  {title} {\bibinfo
  {title} {Room-temperature spin-orbit torque switching induced by a
  topological insulator},\ }\href
  {https://doi.org/10.1103/PhysRevLett.119.077702} {\bibfield  {journal}
  {\bibinfo  {journal} {Phys. Rev. Lett.}\ }\textbf {\bibinfo {volume} {119}},\
  \bibinfo {pages} {077702} (\bibinfo {year} {2017})}\BibitemShut {NoStop}%
\bibitem [{\citenamefont {Huang}\ \emph {et~al.}(2023)\citenamefont {Huang},
  \citenamefont {Islam}, \citenamefont {Ou}, \citenamefont {Ghosh},
  \citenamefont {Richardella}, \citenamefont {Mkhoyan},\ and\ \citenamefont
  {Samarth}}]{Huang_2023}%
  \BibitemOpen
  \bibfield  {author} {\bibinfo {author} {\bibfnamefont {Y.-S.}\ \bibnamefont
  {Huang}}, \bibinfo {author} {\bibfnamefont {S.}~\bibnamefont {Islam}},
  \bibinfo {author} {\bibfnamefont {Y.}~\bibnamefont {Ou}}, \bibinfo {author}
  {\bibfnamefont {S.}~\bibnamefont {Ghosh}}, \bibinfo {author} {\bibfnamefont
  {A.}~\bibnamefont {Richardella}}, \bibinfo {author} {\bibfnamefont {K.~A.}\
  \bibnamefont {Mkhoyan}},\ and\ \bibinfo {author} {\bibfnamefont
  {N.}~\bibnamefont {Samarth}},\ }\bibfield  {title} {\bibinfo {title}
  {Epitaxial growth and characterization of {\BiSb} thin films on (0001)
  sapphire},\ }\href@noop {} {\bibfield  {journal} {\bibinfo  {journal}
  {arXiv}\ ,\ \bibinfo {pages} {xxxx}} (\bibinfo {year} {2023})}\BibitemShut
  {NoStop}%
\bibitem [{\citenamefont {Hsieh}\ \emph {et~al.}(2008)\citenamefont {Hsieh},
  \citenamefont {Qian}, \citenamefont {Wray}, \citenamefont {Xia},
  \citenamefont {Hor}, \citenamefont {Cava},\ and\ \citenamefont
  {Hasan}}]{Hsieh2008}%
  \BibitemOpen
  \bibfield  {author} {\bibinfo {author} {\bibfnamefont {D.}~\bibnamefont
  {Hsieh}}, \bibinfo {author} {\bibfnamefont {D.}~\bibnamefont {Qian}},
  \bibinfo {author} {\bibfnamefont {L.}~\bibnamefont {Wray}}, \bibinfo {author}
  {\bibfnamefont {Y.}~\bibnamefont {Xia}}, \bibinfo {author} {\bibfnamefont
  {Y.~S.}\ \bibnamefont {Hor}}, \bibinfo {author} {\bibfnamefont {R.~J.}\
  \bibnamefont {Cava}},\ and\ \bibinfo {author} {\bibfnamefont {M.~Z.}\
  \bibnamefont {Hasan}},\ }\bibfield  {title} {\bibinfo {title} {{A topological
  Dirac insulator in a quantum spin Hall phase}},\ }\href
  {https://www.nature.com/articles/nature06843} {\bibfield  {journal} {\bibinfo
   {journal} {Nature (London)}\ }\textbf {\bibinfo {volume} {452}},\ \bibinfo
  {pages} {970} (\bibinfo {year} {2008})}\BibitemShut {NoStop}%
\bibitem [{\citenamefont {Benia}\ \emph {et~al.}(2015)\citenamefont {Benia},
  \citenamefont {Stra{\ss}er}, \citenamefont {Kern},\ and\ \citenamefont
  {Ast}}]{Benia2015}%
  \BibitemOpen
  \bibfield  {author} {\bibinfo {author} {\bibfnamefont {H.~M.}\ \bibnamefont
  {Benia}}, \bibinfo {author} {\bibfnamefont {C.}~\bibnamefont {Stra{\ss}er}},
  \bibinfo {author} {\bibfnamefont {K.}~\bibnamefont {Kern}},\ and\ \bibinfo
  {author} {\bibfnamefont {C.~R.}\ \bibnamefont {Ast}},\ }\bibfield  {title}
  {\bibinfo {title} {Surface band structure of {Bi}$_{1-x}${Sb}$_x$(111)},\
  }\href {https://doi.org/10.1103/PhysRevB.91.161406} {\bibfield  {journal}
  {\bibinfo  {journal} {Phys. Rev. B}\ }\textbf {\bibinfo {volume} {91}},\
  \bibinfo {pages} {161406(R)} (\bibinfo {year} {2015})}\BibitemShut {NoStop}%
\bibitem [{\citenamefont {Zhang}\ \emph {et~al.}(2009)\citenamefont {Zhang},
  \citenamefont {Liu}, \citenamefont {Qi}, \citenamefont {Deng}, \citenamefont
  {Dai}, \citenamefont {Zhang},\ and\ \citenamefont
  {Fang}}]{Zhang_PhysRevB.80.085307}%
  \BibitemOpen
  \bibfield  {author} {\bibinfo {author} {\bibfnamefont {H.-J.}\ \bibnamefont
  {Zhang}}, \bibinfo {author} {\bibfnamefont {C.-X.}\ \bibnamefont {Liu}},
  \bibinfo {author} {\bibfnamefont {X.-L.}\ \bibnamefont {Qi}}, \bibinfo
  {author} {\bibfnamefont {X.-Y.}\ \bibnamefont {Deng}}, \bibinfo {author}
  {\bibfnamefont {X.}~\bibnamefont {Dai}}, \bibinfo {author} {\bibfnamefont
  {S.-C.}\ \bibnamefont {Zhang}},\ and\ \bibinfo {author} {\bibfnamefont
  {Z.}~\bibnamefont {Fang}},\ }\bibfield  {title} {\bibinfo {title} {Electronic
  structures and surface states of the topological insulator
  ${\text{bi}}_{1\ensuremath{-}x}{\text{sb}}_{x}$},\ }\href
  {https://doi.org/10.1103/PhysRevB.80.085307} {\bibfield  {journal} {\bibinfo
  {journal} {Phys. Rev. B}\ }\textbf {\bibinfo {volume} {80}},\ \bibinfo
  {pages} {085307} (\bibinfo {year} {2009})}\BibitemShut {NoStop}%
\bibitem [{\citenamefont {Ast}\ and\ \citenamefont
  {H\"ochst}(2003)}]{Ast_PhysRevB.67.113102}%
  \BibitemOpen
  \bibfield  {author} {\bibinfo {author} {\bibfnamefont {C.~R.}\ \bibnamefont
  {Ast}}\ and\ \bibinfo {author} {\bibfnamefont {H.}~\bibnamefont {H\"ochst}},\
  }\bibfield  {title} {\bibinfo {title} {Electronic structure of a bismuth
  bilayer},\ }\href {https://doi.org/10.1103/PhysRevB.67.113102} {\bibfield
  {journal} {\bibinfo  {journal} {Phys. Rev. B}\ }\textbf {\bibinfo {volume}
  {67}},\ \bibinfo {pages} {113102} (\bibinfo {year} {2003})}\BibitemShut
  {NoStop}%
\bibitem [{\citenamefont {Zhang}\ \emph {et~al.}(2016)\citenamefont {Zhang},
  \citenamefont {Velev}, \citenamefont {Dang},\ and\ \citenamefont
  {Tsymbal}}]{zhang2016band}%
  \BibitemOpen
  \bibfield  {author} {\bibinfo {author} {\bibfnamefont {J.}~\bibnamefont
  {Zhang}}, \bibinfo {author} {\bibfnamefont {J.~P.}\ \bibnamefont {Velev}},
  \bibinfo {author} {\bibfnamefont {X.}~\bibnamefont {Dang}},\ and\ \bibinfo
  {author} {\bibfnamefont {E.~Y.}\ \bibnamefont {Tsymbal}},\ }\bibfield
  {title} {\bibinfo {title} {Band structure and spin texture of
  {B}i$_2${S}e$_3$ 3d ferromagnetic metal interface},\ }\href
  {https://journals.aps.org/prb/abstract/10.1103/PhysRevB.94.014435} {\bibfield
   {journal} {\bibinfo  {journal} {Phys. Rev. B}\ }\textbf {\bibinfo {volume}
  {94}},\ \bibinfo {pages} {014435} (\bibinfo {year} {2016})}\BibitemShut
  {NoStop}%
\bibitem [{\citenamefont {Pai}\ \emph {et~al.}(2015)\citenamefont {Pai},
  \citenamefont {Ou}, \citenamefont {Vilela-Le\~ao}, \citenamefont {Ralph},\
  and\ \citenamefont {Buhrman}}]{Pai_PhysRevB.92.064426}%
  \BibitemOpen
  \bibfield  {author} {\bibinfo {author} {\bibfnamefont {C.-F.}\ \bibnamefont
  {Pai}}, \bibinfo {author} {\bibfnamefont {Y.}~\bibnamefont {Ou}}, \bibinfo
  {author} {\bibfnamefont {L.~H.}\ \bibnamefont {Vilela-Le\~ao}}, \bibinfo
  {author} {\bibfnamefont {D.~C.}\ \bibnamefont {Ralph}},\ and\ \bibinfo
  {author} {\bibfnamefont {R.~A.}\ \bibnamefont {Buhrman}},\ }\bibfield
  {title} {\bibinfo {title} {Dependence of the efficiency of spin {H}all torque
  on the transparency of pt/ferromagnetic layer interfaces},\ }\href
  {https://doi.org/10.1103/PhysRevB.92.064426} {\bibfield  {journal} {\bibinfo
  {journal} {Phys. Rev. B}\ }\textbf {\bibinfo {volume} {92}},\ \bibinfo
  {pages} {064426} (\bibinfo {year} {2015})}\BibitemShut {NoStop}%
\bibitem [{\citenamefont {Ou}\ \emph {et~al.}(2016)\citenamefont {Ou},
  \citenamefont {Pai}, \citenamefont {Shi}, \citenamefont {Ralph},\ and\
  \citenamefont {Buhrman}}]{Ou_PhysRevB.94.140414}%
  \BibitemOpen
  \bibfield  {author} {\bibinfo {author} {\bibfnamefont {Y.}~\bibnamefont
  {Ou}}, \bibinfo {author} {\bibfnamefont {C.-F.}\ \bibnamefont {Pai}},
  \bibinfo {author} {\bibfnamefont {S.}~\bibnamefont {Shi}}, \bibinfo {author}
  {\bibfnamefont {D.~C.}\ \bibnamefont {Ralph}},\ and\ \bibinfo {author}
  {\bibfnamefont {R.~A.}\ \bibnamefont {Buhrman}},\ }\bibfield  {title}
  {\bibinfo {title} {Origin of fieldlike spin-orbit torques in heavy
  metal/ferromagnet/oxide thin film heterostructures},\ }\href
  {https://doi.org/10.1103/PhysRevB.94.140414} {\bibfield  {journal} {\bibinfo
  {journal} {Phys. Rev. B}\ }\textbf {\bibinfo {volume} {94}},\ \bibinfo
  {pages} {140414(R)} (\bibinfo {year} {2016})}\BibitemShut {NoStop}%
\bibitem [{\citenamefont {An}\ \emph {et~al.}(2016)\citenamefont {An},
  \citenamefont {Kageyama}, \citenamefont {Kanno}, \citenamefont {Enishi},\
  and\ \citenamefont {Ando}}]{An2016CuOx}%
  \BibitemOpen
  \bibfield  {author} {\bibinfo {author} {\bibfnamefont {H.}~\bibnamefont
  {An}}, \bibinfo {author} {\bibfnamefont {Y.}~\bibnamefont {Kageyama}},
  \bibinfo {author} {\bibfnamefont {Y.}~\bibnamefont {Kanno}}, \bibinfo
  {author} {\bibfnamefont {N.}~\bibnamefont {Enishi}},\ and\ \bibinfo {author}
  {\bibfnamefont {K.}~\bibnamefont {Ando}},\ }\bibfield  {title} {\bibinfo
  {title} {Spin--torque generator engineered by natural oxidation of
  $\mathrm{Cu}$},\ }\href {https://doi.org/10.1038/ncomms13069} {\bibfield
  {journal} {\bibinfo  {journal} {Nat. Commun.}\ }\textbf {\bibinfo {volume}
  {7}},\ \bibinfo {pages} {13069} (\bibinfo {year} {2016})}\BibitemShut
  {NoStop}%
\bibitem [{\citenamefont {Pai}\ \emph {et~al.}(2012)\citenamefont {Pai},
  \citenamefont {Liu}, \citenamefont {Li}, \citenamefont {Tseng}, \citenamefont
  {Ralph},\ and\ \citenamefont {Buhrman}}]{Pai2012}%
  \BibitemOpen
  \bibfield  {author} {\bibinfo {author} {\bibfnamefont {C.~F.}\ \bibnamefont
  {Pai}}, \bibinfo {author} {\bibfnamefont {L.}~\bibnamefont {Liu}}, \bibinfo
  {author} {\bibfnamefont {Y.}~\bibnamefont {Li}}, \bibinfo {author}
  {\bibfnamefont {H.~W.}\ \bibnamefont {Tseng}}, \bibinfo {author}
  {\bibfnamefont {D.~C.}\ \bibnamefont {Ralph}},\ and\ \bibinfo {author}
  {\bibfnamefont {R.~A.}\ \bibnamefont {Buhrman}},\ }\bibfield  {title}
  {\bibinfo {title} {Spin transfer torque devices utilizing the giant spin
  {H}all effect of tungsten},\ }\href {https://doi.org/10.1063/1.4753947}
  {\bibfield  {journal} {\bibinfo  {journal} {Appl. Phys. Lett.}\ }\textbf
  {\bibinfo {volume} {101}},\ \bibinfo {pages} {122404} (\bibinfo {year}
  {2012})}\BibitemShut {NoStop}%
\bibitem [{\citenamefont {Fan}\ \emph {et~al.}(2022)\citenamefont {Fan},
  \citenamefont {Khang}, \citenamefont {Nakano},\ and\ \citenamefont
  {Hai}}]{Fan2022}%
  \BibitemOpen
  \bibfield  {author} {\bibinfo {author} {\bibfnamefont {T.}~\bibnamefont
  {Fan}}, \bibinfo {author} {\bibfnamefont {N.~H.~D.}\ \bibnamefont {Khang}},
  \bibinfo {author} {\bibfnamefont {S.}~\bibnamefont {Nakano}},\ and\ \bibinfo
  {author} {\bibfnamefont {P.~N.}\ \bibnamefont {Hai}},\ }\bibfield  {title}
  {\bibinfo {title} {Ultrahigh efficient spin orbit torque magnetization
  switching in fully sputtered topological insulator and ferromagnet
  multilayers},\ }\href {https://doi.org/10.1038/s41598-022-06779-3} {\bibfield
   {journal} {\bibinfo  {journal} {Sci. Rep.}\ }\textbf {\bibinfo {volume}
  {12}} (\bibinfo {year} {2022})}\BibitemShut {NoStop}%
\bibitem [{\citenamefont {Lee}\ \emph {et~al.}(2014)\citenamefont {Lee},
  \citenamefont {Liu}, \citenamefont {Pai}, \citenamefont {Li}, \citenamefont
  {Tseng}, \citenamefont {Gowtham}, \citenamefont {Park}, \citenamefont
  {Ralph},\ and\ \citenamefont {Buhrman}}]{Lee_PhysRevB.89.024418}%
  \BibitemOpen
  \bibfield  {author} {\bibinfo {author} {\bibfnamefont {O.~J.}\ \bibnamefont
  {Lee}}, \bibinfo {author} {\bibfnamefont {L.~Q.}\ \bibnamefont {Liu}},
  \bibinfo {author} {\bibfnamefont {C.~F.}\ \bibnamefont {Pai}}, \bibinfo
  {author} {\bibfnamefont {Y.}~\bibnamefont {Li}}, \bibinfo {author}
  {\bibfnamefont {H.~W.}\ \bibnamefont {Tseng}}, \bibinfo {author}
  {\bibfnamefont {P.~G.}\ \bibnamefont {Gowtham}}, \bibinfo {author}
  {\bibfnamefont {J.~P.}\ \bibnamefont {Park}}, \bibinfo {author}
  {\bibfnamefont {D.~C.}\ \bibnamefont {Ralph}},\ and\ \bibinfo {author}
  {\bibfnamefont {R.~A.}\ \bibnamefont {Buhrman}},\ }\bibfield  {title}
  {\bibinfo {title} {Central role of domain wall depinning for perpendicular
  magnetization switching driven by spin torque from the spin {H}all effect},\
  }\href {https://doi.org/10.1103/PhysRevB.89.024418} {\bibfield  {journal}
  {\bibinfo  {journal} {Phys. Rev. B}\ }\textbf {\bibinfo {volume} {89}},\
  \bibinfo {pages} {024418} (\bibinfo {year} {2014})}\BibitemShut {NoStop}%
\bibitem [{\citenamefont {Liu}\ and\ \citenamefont {Allen}(1995)}]{LiuAllen}%
  \BibitemOpen
  \bibfield  {author} {\bibinfo {author} {\bibfnamefont {Y.}~\bibnamefont
  {Liu}}\ and\ \bibinfo {author} {\bibfnamefont {R.~E.}\ \bibnamefont
  {Allen}},\ }\bibfield  {title} {\bibinfo {title} {{Electronic structure of
  the semimetals Bi and Sb}},\ }\href
  {https://doi.org/10.1103/PhysRevB.52.1566} {\bibfield  {journal} {\bibinfo
  {journal} {Phys. Rev. B}\ }\textbf {\bibinfo {volume} {52}},\ \bibinfo
  {pages} {1566} (\bibinfo {year} {1995})}\BibitemShut {NoStop}%
\end{thebibliography}
\providecommand{\noopsort}[1]{}\providecommand{\singleletter}[1]{#1}%

\end{document}